\documentclass[twocolumn]{IEEEtran}
\usepackage[T1]{fontenc}
\usepackage[latin9]{inputenc}
\usepackage{color}
\usepackage{array}
\usepackage{float}
\usepackage{textcomp}
\usepackage{amsmath}
\usepackage{amssymb}
\usepackage{graphicx}
\usepackage[unicode=true,
 bookmarks=true,bookmarksnumbered=true,bookmarksopen=true,bookmarksopenlevel=1,
 breaklinks=false,pdfborder={0 0 0},pdfborderstyle={},backref=false,colorlinks=false]
 {hyperref}
\hypersetup{pdftitle={Your Title},
 pdfauthor={Your Name},
 pdfpagelayout=OneColumn, pdfnewwindow=true, pdfstartview=XYZ, plainpages=false}

\makeatletter

\providecommand{\tabularnewline}{\\}
\floatstyle{ruled}
\newfloat{algorithm}{tbp}{loa}
\providecommand{\algorithmname}{Algorithm}
\floatname{algorithm}{\protect\algorithmname}
\usepackage{tikz}
\usetikzlibrary{calc}

\newenvironment{lyxlist}[1]
	{\begin{list}{}
		{\settowidth{\labelwidth}{#1}
		 \setlength{\leftmargin}{\labelwidth}
		 \addtolength{\leftmargin}{\labelsep}
		 }}
	{\end{list}}

\usepackage{cite}

\usepackage{algorithmic}
\usepackage{subfig}

\@ifundefined{showcaptionsetup}{}{%
 \PassOptionsToPackage{caption=false}{subfig}}
\usepackage{subfig}
\makeatother

\begin{document}
\title{A Two-Stage 2D Channel Extrapolation Scheme for TDD 5G NR Systems}
\author{{\normalsize{}Yubo Wan, }\textit{\normalsize{}Graduate Student Member,
IEEE}{\normalsize{}, and An Liu, }\textit{\normalsize{}Senior Member,
IEEE}{\normalsize{}}\thanks{This work was supported by National Key R\&D Program of China (Grant
No. 2021YFA1003304). (Corresponding authors: An Liu.)

Yubo Wan, An Liu are with the College of Information Science and Electronic
Engineering, Zhejiang University, Hangzhou 310027, China (email: wanyb@zju.edu.cn,
anliu@zju.edu.cn).}\vspace{-0.4in}
}
\maketitle
\begin{abstract}
Recently, channel extrapolation has been widely investigated in FDD
massive MIMO systems. However, in TDD  5G new radio (NR) systems,
the channel extrapolation problem also arises due to the hopping uplink
pilot pattern, which has not been fully researched yet. This paper
addresses this gap by formulating a channel extrapolation problem
in TDD massive MIMO-OFDM systems for 5G NR, incorporating imperfection
factors. A novel two-stage 2D channel extrapolation scheme in  frequency-time
domain is proposed, designed to mitigate the  effects of imperfection
factors and ensure high-accuracy channel estimation. Specifically,
in the channel estimation stage, we propose a novel multi-band  multi-timeslot
based high-resolution parameter estimation algorithm to achieve 2D
channel extrapolation in the presence of imperfection factors. Then,
to avoid repeated multi-timeslot  channel estimation, a channel tracking
stage is designed during the subsequent time instants, where a sparse
Markov channel model is formulated to capture the dynamic sparsity
of massive MIMO-OFDM channels under the influence of imperfection
factors. Next, an expectation-maximization (EM) based compressive
channel tracking algorithm is designed to  estimate unknown imperfection
and channel parameters by exploiting the high-resolution prior information
of the delay/angle parameters from  previous timeslots. Simulation
results underscore the superior performance of our proposed channel
extrapolation scheme over baselines.
\end{abstract}


\section{Introduction}

Massive multiple input multiple output (MIMO) presents a viable technology
in fifth-generation (5G) New Radio (NR) systems, capable of utilizing
substantial spatial multiplexing gain to satisfy escalating demands
for large communication capacity \cite{Larsson2014Massive,lu2014overview}.
Yet, the deployment of massive MIMO communication systems hinges on
obtaining precise channel state information (CSI) at the base station
(BS), which is a challenge for a practical time-varying wireless channel
subject to system imperfections.

\begin{figure}[t]
\begin{centering}
\textsf{\includegraphics[width=9cm]{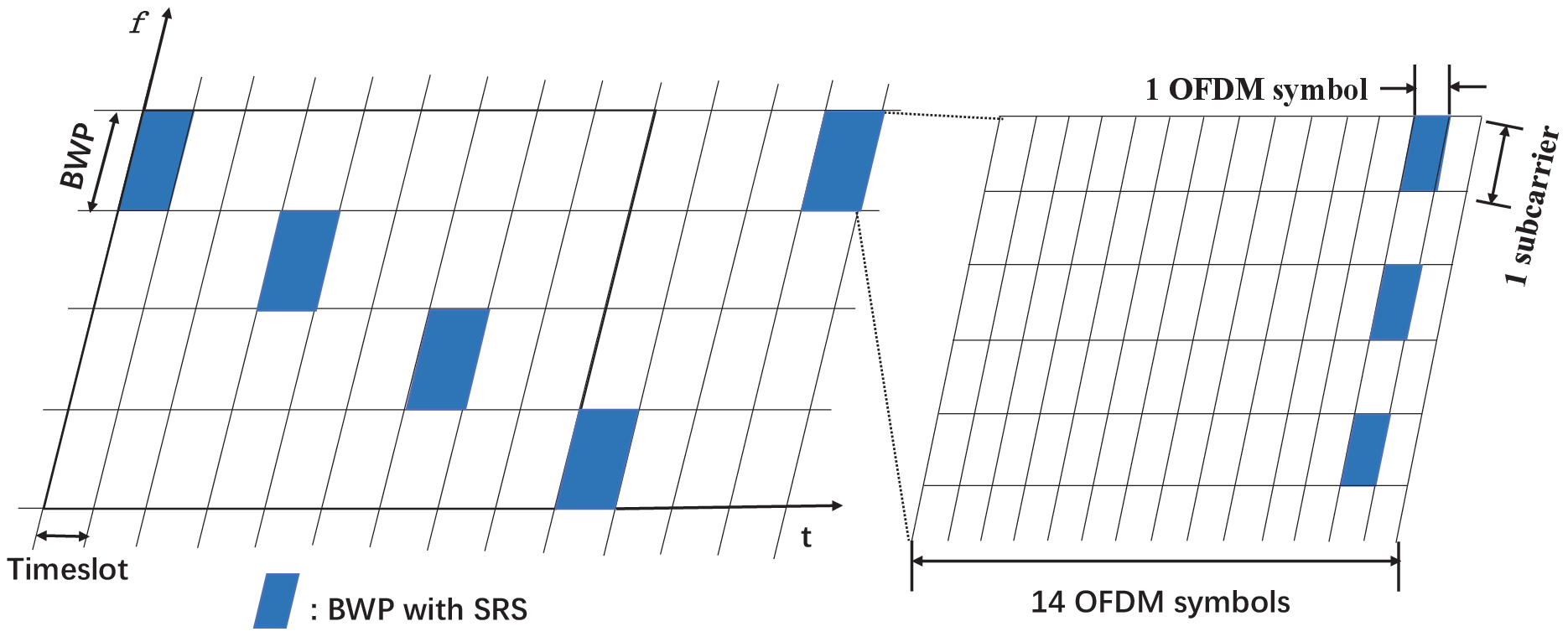}}
\par\end{centering}
\caption{\label{fig:Hopping}An illustration of the hopping SRS pattern.}
\end{figure}

In time division duplex (TDD) massive MIMO systems, the BS can efficiently
estimate downlink channels based on the received uplink pilots (also
called Sounding Reference Signals (SRSs) in 5G NR systems \cite{3gpp_5GNR})
transmitted from the user due to channel reciprocity. As compared
to frequency division duplex (FDD) systems, TDD systems drastically
reduce the pilot overhead for CSI acquisition at the BS. However,
owing to the user's limited transmission power, the user typically
transmits the SRSs within a bandwidth part (BWP), occupying only a
fraction of the entire system bandwidth for a given SRS period \cite{3gpp_5GNR}.
Different timeslots adopt a frequency hopping pattern, depicted in
Fig. \ref{fig:Hopping}, where the blue part denotes the time-frequency
resource block occupied by the SRS symbols, with a total of $h_{p}$
hops, e.g., $h_{p}=4$ in Fig. \ref{fig:Hopping}. This approach ensures
that the power spectral density per subcarrier of the transmitted
SRSs surpasses that of a full-band SRSs transmission scheme, enhancing
the transmission system's anti-noise interference capability. However,
this method  estimates only a subset of the uplink channels in the
time-frequency domain using received SRSs, necessitating the extrapolation
method to deduce the remaining channels.

Numerous channel estimation/extrapolation methodologies have been
proposed in massive MIMO systems. Traditional methods such as the
least square (LS) approach \cite{kalachikov2021performance} and minimum
mean square error (MMSE) method \cite{sengijpta1995fundamentals,Extrapolation_1}
have been widely employed, but their extrapolation capabilities are
limited. Conversely, high-resolution parameter estimation (HRPE) algorithms,
which estimate the multipath components (MPCs), can achieve wide frequency
range channel extrapolation, e.g., subspace-based algorithms \cite{Extrapolation_2_MUSIC,Extrapolation_3_ESPRIT},
compressed sensing (CS) methods \cite{Extrapolation_4_SBL}, and etc.
A significant volume of research has explored channel extrapolation
in the frequency domain of the FDD systems, e.g., \cite{Extrapolation_1,Extrapolation_5,Extrapolation_6_DeepL,Extrapolation_3_ESPRIT},
which extrapolates the uplink channel to the downlink channel by leveraging
their pronounced spatial correlations. This correlation exists because
signals at different frequencies propagate within the same environment
and follow identical propagation paths \cite{FDD_reciprocity}. Consequently,
this approach eliminates the overhead associated with pilot transmission
and feedback for the downlink channel \cite{FDD_feedback}. In \cite{Extrapolation_1},
the authors analyzed the theoretical performance bound of channel
extrapolation in FDD massive MIMO systems. In \cite{Extrapolation_6_DeepL},
machine learning methods have been applied to extrapolate downlink
CSI from observed uplink CSI in MIMO systems. The channel extrapolation
in time domain has also been examined, e.g., \cite{Extrapolation_7_time,Extrapolation_8_time_fre},
and a multi-domain channel extrapolation scheme  has been proposed
in \cite{Extrapolation_9_multi}.

Nevertheless, there are limitations of the existing work: 

(i) Few studies have explored channel extrapolation in TDD massive
MIMO systems, focusing more on FDD systems. However, in FDD systems,
pilots are uniformly assigned across the entire system bandwidth and
additional pilots are usually transmitted in the downlink to estimate
the channel path coefficients since the uplink and downlink channels
in FDD do not share the same channel path coefficients. As a result,
the existing channel extrapolation schemes for FDD systems cannot
work in TDD 5G NR systems that need to estimate both delay/angle parameters
and channel path coefficients from the frequency hopping pilots.

(ii) Most research only conducts one-dimensional channel extrapolation
in frequency or time domain. Moreover, the imperfection factors such
as the Doppler frequency offset, time offset, and random phase noise
in real systems \cite{Doppler,XuHuilin2}, distort the time correlation
 of the channel, which further complicates channel extrapolation in
both  time and frequency domains.

Motivated by the limitations of existing channel extrapolation methods
and the challenges brought by the imperfection factors, this paper
considers a TDD massive MIMO-OFDM scenario, bearing in mind the practical
imperfections in 5G NR systems. We propose a novel two-stage 2D channel
extrapolation scheme that operates in both  frequency and time domain.
This scheme is compatible with 5G NR and can circumvent all imperfections
to achieve high-accuracy channel parameter estimation. The main contributions
of this paper are summarized as follows.
\begin{enumerate}
\item We propose a novel two-stage 2D channel extrapolation scheme, including
a new extrapolation signal model for TDD 5G NR systems and a two-stage
channel extrapolation algorithm. The signal model designed for 2D
channel extrapolation is time-variant and considers a hopping pilot
pattern and multiple imperfection factors in real 5G NR systems. In
the first stage, we perform a multi-band multi-timeslot channel estimation
by combining the received SRS observations from multiple timeslots
and distinct BWPs to realize a 2D channel extrapolation. Subsequently,
for the sake of keeping the CSI fresh and accurate, meanwhile, avoiding
huge complexities caused by frequent multi-timeslots based channel
estimation, we perform channel extrapolation in the following timeslots
using channel tracking methods in the second stage by exploiting high-resolution
prior information of channel parameters passed from the previous timeslot.
\item In the first stage (channel estimation stage), we propose a multi-band
 multi-timeslot HRPE (MBMT-HRPE) scheme based on a novel robust time-space-time
multiple signal classification (R-TST-MUSIC) algorithm, which is an
extension of the TST-MUSIC \cite{TST_MUSIC} from single-band/single-timeslot
to multi-band and multi-timeslot. Unlike traditional single-band or
single-timeslot based MUSIC algorithm, which cannot achieve  HRPE
 due to the effect of imperfection factors, our proposed R-TST-MUSIC
algorithm is able to coherently combine observations from various
timeslots and BWPs to obtain equivalent full-band observations by
compensating for the time-variant imperfection factors, thereby achieving
high-accuracy parameter estimation. Specifically, following algorithm
initialization, the proposed R-TST-MUSIC algorithm performs alternating
optimization (AO) iterations between three components: Multi-band
observations splicing; Joint delay-angle channel parameters estimation
using a TST-MUSIC-SIC method (a combination of the TST-MUSIC and successive
interference cancellation); Imperfection factors and channel coefficients
estimation based on the maximum likelihood (ML) method. Such a MBMT-HRPE
stage is crucial because channel extrapolation has a very high requirement
on the delay estimation accuracy. Therefore, it is necessary for the
initial stage to provide a high-resolution prior information of the
delay parameters for the subsequent tracking stage.
\item In the second stage (channel tracking stage), a sparse Markov channel
model is developed to capture the time correlation and  sparsity of
the massive MIMO-OFDM channels while considering imperfection factors.
Then, a robust channel tracking scheme is proposed to achieve channel
extrapolation at each timeslot based on expectation-maximization (EM)
method. During the E-Step, we employ the dynamic Turbo-CS method and
message passing method to leverage the time correlation  and sparsity
of the channel to accomplish Bayesian channel estimation. Then, in
the M-Step, given the Bayesian channel estimation results and the
high-resolution prior information from the previous timeslot, both
delay/angle off-grid and imperfection parameters are estimated to
further enhance the channel extrapolation accuracy and the robustness
against system imperfections.
\end{enumerate}

The rest of this paper is organized as follows. In Section \ref{sec:System-Model},
we describe the system and signal model. In Section \ref{sec:EST}
and \ref{sec:Tra}, we present the proposed channel extrapolation
scheme in the channel estimation stage and the channel tracking stage,
respectively. Finally, the simulation results and conclusions are
given in Sections \ref{sec:Simulation-Results} and \ref{sec:Conclusion},
respectively. 

\textit{Notations:} The notation $\left\Vert \cdot\right\Vert _{F}$
denotes the Frobenius norm, $\angle(\cdot)$ denotes the phase of
a complex scalar, $vec(\cdot)$ denotes the vectorization, $\mathrm{diag}\left(\cdot\right)$
constructs a diagonal matrix from its vector argument, $\odot$, $\otimes$,
and $\ast$ denote the Khatri-Rao product, Kronecker product, and
Hadamard product, respectively. The transpose, conjugate transpose,
and inverse are denoted by $(\cdot)^{T},(\cdot)^{H},(\cdot)^{-1}$
respectively. $\mathcal{CN}(\mathbf{x};\mathrm{\boldsymbol{\mu}},\boldsymbol{\Sigma})$
denotes a complex Gaussian normal distribution corresponding to variable
$\mathbf{x}$ with mean $\boldsymbol{\mu}$ and covariance matrix
$\boldsymbol{\Sigma}$.\vspace{-0.2in}

\section{System and Signal Model\label{sec:System-Model}}

\subsection{System Model}

We consider a TDD massive MIMO-OFDM system, where each single antenna
user transmits uplink hopping SRSs and moves with low speed. The SRSs
have been specified in the 3GPP 5G NR Release 16, which is obtained
from Zadoff-Chu (ZC) sequence \cite{3gpp_5GNR}. Note that the extension
of our scenario to a user with multiple antennas is trivial, since
the antennas in the same user are generally assigned with orthogonal
SRSs. Moreover, since different users in the same cell transmit SRSs
at different subcarriers/OFDM symbols, we focus on the channel extrapolation
problem for a single user in this paper. The BS is equipped with $N_{r}=N_{x}\times N_{y}$
uniform planar array (UPA) antennas, where $N_{x}$ and $N_{y}$ represent
the antenna number in the horizontal and vertical direction, respectively.

In 5G NR systems,  a timeslot  contains 14 OFDM symbols. To save pilot
overhead and improve the spectrum efficiency, the SRS is periodically
transmitted at the consecutive 1, 2 or 4 OFDM symbols in a timeslot
with the period of $T_{S}$ timeslots \cite{3gpp_5GNR}. In  frequency
domain, the SRS pattern for a user has a comb structure in its allocated
BWP, i.e., the SRS is transmitted on every $N_{c}$ subcarrier in
the BWP. Without loss of generality, we focus on the comb-2 mode in
this paper, i.e., $N_{c}=2$, and the SRS symbols are transmitted
at 1 OFDM symbol in a timeslot, as shown in Fig. \ref{fig:Hopping}.

\vspace{-0.1in}

\subsection{Signal Model}

In the $t$-th SRS symbol, the channel frequency response (CFR) at
the $n$-th ($0\leq n\leq N-1$) subcarrier is given by
\begin{flalign}
\mathbf{h}^{(t)}\left[n\right] & =e^{j\varepsilon^{(t)}}\!\!\sum_{k=1}^{K^{(t)}}\!\alpha_{k}^{(t)}e^{j\varphi_{k}^{(t)}}\!e^{-j2\pi nf_{s}\left(\tau_{k}^{(t)}+\tau_{0}^{(t)}\right)}\!\boldsymbol{a}_{R}\!\left(\!\theta_{k}^{(t)},\phi_{k}^{(t)}\!\right),\label{eq:H channel model F}
\end{flalign}
where $K^{(t)}$ is the number of propagation paths, $f_{s}$ denotes
the subcarrier spacing, $\alpha_{k}^{(t)}$, $\theta_{k}^{(t)}$,
$\phi_{k}^{(t)}$, and $\tau_{k}^{(t)}$ denote the complex path gain,
azimuth angle of arrival (AoA), elevation AoA and time delay of the
$k$-th path, respectively. The notations $\varphi_{k}^{(t)}$ , $\varepsilon^{(t)}$,
and $\tau_{0}^{(t)}$ denote the imperfection factors, i.e., Doppler
phase rotation factor caused by user mobility, random phase noise,
and the time offset \cite{Doppler,XuHuilin2}. Without loss of generality,
we set $\varepsilon^{(1)}=0,\tau_{0}^{(1)}=0,\varphi_{k}^{(1)}=0,\forall k,$
for the first SRS symbol. The factor $\varphi_{k}^{(t)}$ depends
on the Doppler frequency offset $f_{D,k}^{(t)}$ as $\varphi_{k}^{(t)}=\varphi_{k}^{(t-1)}+2\pi T_{SRS}f_{D,k}^{(t)}$,
where $T_{SRS}$ is the time interval between two adjacent SRS symbols.
Since we focus on the scenario with a low-speed and acceleration of
the user, $f_{D,k}^{(t)}$ varies slowly and has a strong time-correlation
to be exploited. $\boldsymbol{a}_{R}\left(\theta,\phi\right)=\boldsymbol{a}_{x}\left(\theta,\phi\right)\otimes\boldsymbol{a}_{y}\left(\theta\right)\in\mathbb{C}^{N_{r}\times1}$
is the array response vector for the BS antenna array. With half-wavelength
spacing, the $n_{x}$-th element in the steering vector $\boldsymbol{a}_{x}\left(\theta,\phi\right)$
and the $n_{y}$-th element in the steering vector $\boldsymbol{a}_{y}\left(\theta\right)$
can be respectively expressed as $\left[\boldsymbol{a}_{x}\left(\theta,\phi\right)\right]_{n_{x}}=\frac{1}{\sqrt{N_{x}}}e^{j\pi n_{x}sin\left(\theta\right)cos\left(\phi\right)},\left[\boldsymbol{a}_{y}\left(\theta\right)\right]_{n_{y}}=\frac{1}{\sqrt{N_{y}}}e^{j\pi n_{y}cos\left(\theta\right)},$
with $n_{x}=0,...,N_{x}-1$ and $n_{y}=0,...,N_{y}-1$ \cite{UPA}.

Let $P$ denote the number of SRS subcarriers in a BWP. Then, the
received frequency domain SRSs $\mathbf{Y}^{(t)}\in\mathbb{C}^{P\text{\texttimes}N_{r}}$
in a BWP for the $t$-th SRS symbol is given by
\begin{alignat}{1}
\mathbf{Y}^{(t)} & =\textrm{diag}\left(\mathbf{x}\right)\mathbf{W}^{(t)}\mathbf{H}^{(t)}+\mathbf{N}^{(t)},\label{eq:recF}
\end{alignat}
where $\mathbf{x}\in\mathbb{C}^{P\text{\texttimes}1}$ is the frequency
domain SRSs transmitted from the user, $\mathbf{W}^{(t)}\in\left\{ 0,1\right\} ^{P\times N}$
is a selection matrix depending on the hopping SRS pattern, e.g.,
$\mathrm{W^{(1)}}\left(i,j\right)=1$ for $j=2i-1$, $i=1,...,P$,
$\mathbf{H}^{(t)}\triangleq[(\mathbf{h}^{(t)}[0])^{T};(\mathbf{h}^{(t)}[1])^{T};...;(\mathbf{h}^{(t)}[N-1])^{T}]\in\mathbb{C}^{N\times N_{r}}$
denotes the CFR matrix on the full-band, and $\mathbf{N}^{\left(t\right)}$
is the additive white Gaussian noise (AWGN) with each element having
zero mean and variance $\sigma_{e}^{(t)}$.

\section{Channel Extrapolation in Channel Estimation Stage\label{sec:EST}}

\begin{figure}[t]
\begin{centering}
\textsf{\includegraphics[width=7cm]{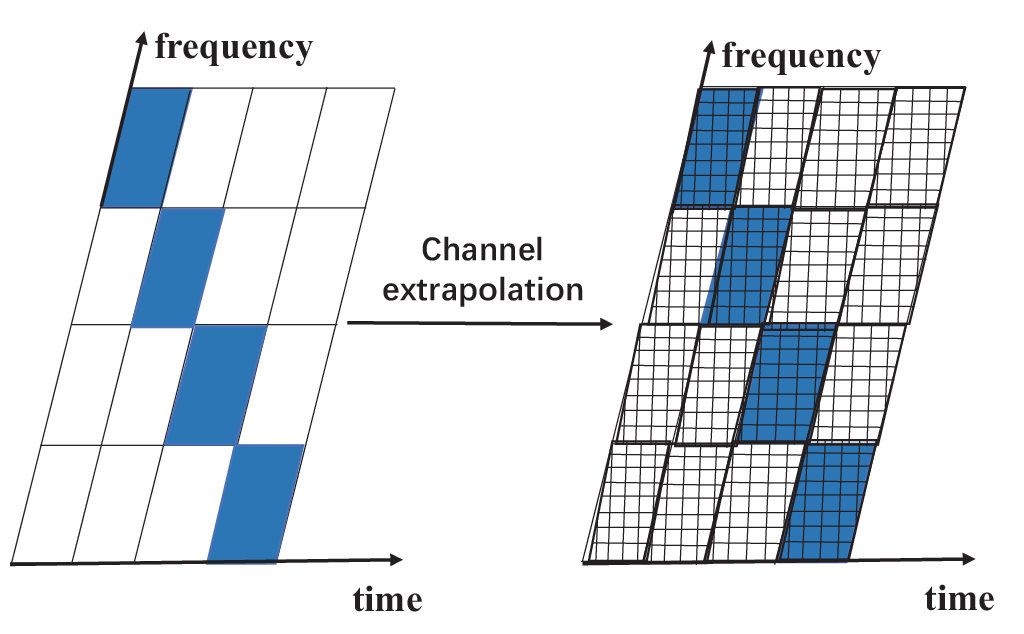}}
\par\end{centering}
\caption{\label{fig:extrapolation_est}Channel extrapolation in channel estimation
stage.}
\end{figure}

In this section, we aim to achieve a 2D channel extrapolation based
on the received hopping SRSs, as illustrated in Fig. \ref{fig:extrapolation_est}.
We first reformulate the received signal model and give an overview
of the TST-MUSIC algorithm. Nevertheless, the original TST-MUSIC algorithm
cannot use the multi-timeslot multi-band observations in a coherent
way due to the effect of imperfection factors, which motivates the
proposed R-TST-MUSIC algorithm.
\vspace{-0.15in}

\subsection{Review of TST-MUSIC Algorithm}

we reformulate the received signal model (\ref{eq:recF}) as:

\begin{equation}
\mathbf{Y}^{(t)}=\mathbf{G}^{(t)}\left(\boldsymbol{\tau}+\tau_{0}^{(t)}\right)\mathbf{B}\mathbf{A}\left(\theta,\phi\right)^{T}+\mathbf{N}^{(t)},\label{eq:MUSIC_Signal}
\end{equation}
where $\mathbf{A}\left(\theta,\phi\right)=[\boldsymbol{a}_{R}(\theta_{1}^{(t)},\phi_{1}^{(t)}),...,\boldsymbol{a}_{R}(\theta_{K^{(t)}}^{(t)},\phi_{K^{(t)}}^{(t)})]\in\mathbb{C}^{N_{r}\times K^{(t)}}$,
$\mathbf{B}=\textrm{diag}(\widetilde{\alpha}_{1}^{(t)},...,\widetilde{\alpha}_{K^{(t)}}^{(t)})$
with $\widetilde{\alpha}_{k}^{(t)}=\alpha_{k}^{(t)}e^{j\varphi_{k}^{(t)}}e^{j\varepsilon^{(t)}}$,
$\mathbf{G}^{(t)}(\boldsymbol{\tau}+\tau_{0}^{(t)})=\textrm{diag}(\mathbf{x})\mathbf{W}^{(t)}\mathbf{F}(\boldsymbol{\tau}+\tau_{0}^{(t)})\triangleq[\mathbf{g}^{(t)}(\tau_{1}^{(t)}+\tau_{0}^{(t)}),...,\mathbf{g}^{(t)}(\tau_{K^{(t)}}^{(t)}+\tau_{0}^{(t)})]\in\mathbb{C}^{P\times K^{(t)}},\mathbf{F}(\boldsymbol{\tau}+\tau_{0}^{(t)})=[\mathbf{a}(\tau_{1}^{(t)}+\tau_{0}^{(t)}),...,\mathbf{a}(\tau_{K^{(t)}}^{(t)}+\tau_{0}^{(t)})]$
with $\mathbf{a}(\tau)=[1,\ldots,e^{-j2\pi(N-1)f_{s}\tau}]^{T}\in\mathbb{C}^{N\times1}$.
Since the user moves with a low speed and the channel varies slowly
as we considered, we mildly assume that the channel parameters are
time-invariant during a small number of initial timeslots in the channel
estimation stage (e.g., for $h_{p}=4,$ we set the number of timeslots
of the channel estimation stage $T_{e}=h_{p}=4$), i.e., $f_{D,k}^{(t)}=f_{D,k},$
$\varphi_{k}^{(t)}=(t-1)\varphi_{k}$, $K^{(t)}=K,\tau_{k}^{(t)}=\tau_{k},\alpha_{k}^{(t)}=\alpha_{k},\theta_{k}^{(t)}=\theta_{k},\phi_{k}^{(t)}=\phi_{k},\forall t\in\left\{ 1,...,T_{e}\right\} $.
But in the simulations, the channel observations are still generated
from a practical channel model specified by 3GPP TR 38.901 without
adding this assumption to fairly evaluate our proposed channel extrapolation
scheme. Simulation results show that our proposed channel extrapolation
scheme works well in practical scenarios though we make such an assumption
in our algorithm design, as detailed in Section \ref{sec:Simulation-Results}.

Then, we give an overview of the TST-MUSIC algorithm, which combines
T-MUSIC and S-MUSIC algorithms along with the temporal filtering techniques
and the spatial beamforming techniques to jointly estimate the angles
and the delays of the multipaths in a wireless channel \cite{TST_MUSIC}.
The TST-MUSIC algorithm has the advantages of high-resolution, which
can resolve paths with either very close angles or very close delays
and automatically pair the estimated angles and delays. For conciseness,
we omit the superscript $(t)$ and focus on a single timeslot to depict
the TST-MUSIC algorithms.

Specifically, the angles and delays are estimated by the S-MUSIC and
T-MUSIC algorithms, which use the covariance matrices of the rows
and the columns of $\mathbf{Y}$, respectively. We perform eigendecomposition
of the autocorrelation matrix of (\ref{eq:MUSIC_Signal}) as
\begin{eqnarray}
\mathbf{R}^{d} & = & \mathbb{E}\left\{ \mathbf{Y}\mathbf{Y}^{H}\right\} =\mathbf{V}_{s}^{d}\boldsymbol{\Lambda}_{s}^{d}\mathbf{V}_{s}^{d\thinspace H}+\mathbf{V}_{n}^{d}\boldsymbol{\Lambda}_{n}^{d}\mathbf{V}_{n}^{d\thinspace H},\label{eq:eigendecom_t}\\
\mathbf{R}^{s} & = & \mathbb{E}\left\{ \mathbf{Y}^{T}\mathbf{Y}^{*}\right\} =\mathbf{V}_{s}^{s}\boldsymbol{\Lambda}_{s}^{s}\mathbf{V}_{s}^{s\thinspace H}+\mathbf{V}_{n}^{s}\boldsymbol{\Lambda}_{n}^{s}\mathbf{V}_{n}^{s\thinspace H},\label{eq:eigendecom_s}
\end{eqnarray}
where the column vectors of $\mathbf{V}_{s}^{d}$ and $\mathbf{V}_{s}^{s}$
are the eigenvectors that span the signal subspace of $\mathbf{R}^{d}$
and $\mathbf{R}^{s}$, respectively, corresponding to the largest
$K$ eigenvalues. The number of multipaths $K$ can be estimated using
the minimum descriptive length (MDL) criterion \cite{MDL}. And the
column vectors of $\mathbf{V}_{n}^{d}$ and $\mathbf{V}_{n}^{s}$
are the eigenvectors that span the noise subspace of $\mathbf{R}^{d}$
and $\mathbf{R}^{s}$, respectively. $\boldsymbol{\Lambda}_{s}^{d},\boldsymbol{\Lambda}_{n}^{d},\boldsymbol{\Lambda}_{s}^{s},\boldsymbol{\Lambda}_{n}^{s}$
are diagonal matrices consisting of the associated eigenvalues. Then,
according to the orthogonality property between the signal and the
noise subspace given by \cite{MUSICbase}
\begin{eqnarray}
\mathbf{G}^{H}\mathbf{V}_{n}^{d} & = & \boldsymbol{O},\label{eq:orthogonality_delay}\\
\mathbf{A}^{H}\mathbf{V}_{n}^{s} & = & \boldsymbol{O},
\end{eqnarray}
i.e., $\mathbf{g}(\tau_{k}+\tau_{0})^{H}\mathbf{V}_{n}^{d}=\mathbf{0}^{T},$$\boldsymbol{a}_{R}(\theta_{k},\phi_{k})^{H}\mathbf{V}_{n}^{s}=\mathbf{0}^{T},\forall k,$
the delays and angles can be estimated at which the following T-MUSIC
and S-MUSIC pseudospectrums achieve maximum values, respectively:
\begin{eqnarray}
\mathcal{P}_{\text{}}^{d}(\tau) & = & \frac{1}{\mathbf{g}(\tau)^{H}\left(\mathbf{I}-\mathbf{V}_{s}^{d}\mathbf{V}_{s}^{d\thinspace H}\right)\mathbf{g}(\tau)},\label{eq:spectrium_t}\\
\mathcal{P}_{\text{}}^{s}(\theta,\phi) & = & \frac{1}{\boldsymbol{a}_{R}(\theta,\phi)^{H}\left(\mathbf{I}-\mathbf{V}_{s}^{s}\mathbf{V}_{s}^{s\thinspace H}\right)\boldsymbol{a}_{R}(\theta,\phi)}.\label{eq:spectrium_s}
\end{eqnarray}

In summary, the TST-MUSIC algorithm has five steps:

Step 1) \textbf{Grouping}: Apply the T-MUSIC algorithm to obtain the
group delays $\left\{ \hat{t}_{1},...,\hat{t}_{q}\right\} $ based
on (\ref{eq:eigendecom_t}) and (\ref{eq:spectrium_t}).

Step 2) \textbf{Temporal Filtering}: The output of the $k$-th group
after filtering is given by
\begin{equation}
\mathbf{Y}_{k}=\prod_{n=1;n\neq k}^{q}\mathbf{U}_{n}^{d}\cdot\mathbf{Y},k=1,...,q,
\end{equation}
where $\mathbf{U}_{n}^{d}=\mathbf{I}-\frac{1}{P}\mathbf{g}\left(\hat{t}_{n}\right)\mathbf{g}^{H}\left(\hat{t}_{n}\right)$
are the temporal filtering matrices.

Step 3) \textbf{DOA Estimation}: Apply the S-MUSIC algorithm to each
$\mathbf{Y}_{k}$ and estimate the angles at each group given by
\begin{equation}
\left(\hat{\boldsymbol{\theta}}_{k},\hat{\boldsymbol{\phi}}_{k}\right)=\left[\left(\hat{\theta}_{k,1},\hat{\phi}_{k,1}\right),...,\left(\hat{\theta}_{k,r(k)},\hat{\phi}_{k,r(k)}\right)\right]^{T},
\end{equation}
where $r(k)$ is the number of paths in the $k$-th group.

Step 4) \textbf{Spatial Beamforming}: The output of the $m$-th spatial
beamformer is given by
\begin{equation}
\mathbf{Y}_{k,m}=\mathbf{Y}_{k}\cdot\prod_{n=1;n\neq m}^{r(k)}\mathbf{U}_{k,n}^{s},k=1,...,K,m=1,...,r(k)
\end{equation}
where $\mathbf{U}_{k,n}^{s}=\mathbf{I}-\boldsymbol{a}_{R}\left(\hat{\theta}_{k,n},\hat{\phi}_{k,n}\right)\boldsymbol{a}_{R}^{H}\left(\hat{\theta}_{k,n},\hat{\phi}_{k,n}\right)$
are the spatial beamforming matrix.

Step 5) \textbf{Delay Estimation}: We again employ the T-MUSIC algorithm
for each $\mathbf{Y}_{k,m}$ to obtain the corresponding delays.\vspace{-0.1in}

\subsection{Outline of the R-TST-MUSIC Algorithm}

It is evident that TST-MUSIC cannot coherently use the BWP observations
across different timeslots to perform channel extrapolation with a
high-accuracy delay estimation. This limitation stems from its inability
to compensate for time-variant imperfection factors. As a result,
when only a fraction of observations is available at each timeslot
(e.g., a quarter of the full-band observations for $h_{p}=4$), the
channel extrapolation capability of TST-MUSIC is limited. However,
using the multi-timeslot multi-band observations in a coherent way
to achieve a high-accuracy channel extrapolation is challenging due
to the following reasons: (i) The imperfection factors destroy the
coherence property of the received BWP observations and thus require
compensation; (ii) The original orthogonality in (\ref{eq:orthogonality_delay})
is affected by the Doppler factors, complicating the estimation of
multipath delay parameters, as elaborated in Subsection \ref{subsec:Step-2}.

To overcome aforementioned challenges, we propose the R-TST-MUSIC
algorithm, an enhancement of the TST-MUSIC algorithm, allowing for
coherent utilization of multi-timeslot multi-band observations. Compared
to the original TST-MUSIC algorithm,  R-TST-MUSIC  is designed to
produce equivalent full-band observations by compensating for the
imperfection factors, yielding superior channel extrapolation performance.
The proposed R-TST-MUSIC algorithm comprises two stages: an initialization
phase and a refinement estimation phase. The former focuses on the
preliminary parameter estimation, utilizing the TST-MUSIC and LS algorithms.
This initial process sets the stage for the refinement phase, ensuring
convergence to a good solution. Armed with the preliminary findings,
the refinement phase then seeks a more precise solution.

The overall R-TST-MUSIC algorithm is summarized in Algorithm \ref{alg:R-TST-MUSIC}.

\begin{algorithm}[t]
\caption{\label{alg:R-TST-MUSIC}R-TST-MUSIC algorithm}

\textbf{Input:} $\mathbf{Y}^{(t)}$, AO iteration number $I_{AO}$.

\textbf{Output:} $\hat{\tau}_{k},\hat{\theta}_{k},\hat{\phi}_{k},\hat{\varphi}_{k},\hat{\varepsilon}^{(t)},\hat{\tau}_{0}^{(t)},t=1,...,T_{e}$.

\begin{algorithmic}[1]

\STATE\textbf{Initialization Phase:}

\STATE Perform TST-MUSIC algorithm based on $\mathbf{Y}^{(t)}$ to
obtain the estimate $\hat{\tau}_{k}^{(t)},\hat{\theta}_{k}^{(t)},\hat{\phi}_{k}^{(t)},\forall t$.

\STATE Get a LS solution $\hat{\widetilde{\boldsymbol{\alpha}}}^{(t)}$in
(\ref{eq:coarse_alpha}).

\STATE Get the estimate $\hat{\varepsilon}^{(t)},\hat{\varphi}_{k}^{(t)},\hat{\tau}_{0}^{(t)}$
in (\ref{eq:coarse_fy})-(\ref{eq:Coarse_tau0}).

\STATE \textbf{Refinement Estimation Phase:}

\FOR{${\color{blue}{\color{black}j=1,\cdots,I_{AO}}}$}

\STATE Construct a compensated signal model (\ref{eq:compensated}).

\STATE Perform TST-MUSIC-SIC algorithm to estimate delay and angle
parameters denoted as $\hat{\tau}_{k},\hat{\theta}_{k},\hat{\phi}_{k}$.

\STATE Get the estimate $\hat{\boldsymbol{\alpha}}^{(j+1)}$,  $\hat{\varepsilon}^{(t)\thinspace(j+1)},\boldsymbol{\varphi}^{(j+1)},\hat{\tau}_{0}^{(t)\thinspace(j+1)}$
based on (\ref{eq:refine_alpha})-(\ref{eq:updaterule_tau}).

\ENDFOR

\end{algorithmic}
\end{algorithm}
\vspace{-0.1in}

\subsection{Initialization  Phase}

At the $t$-th SRS symbol with input $\mathbf{Y}^{(t)}$, we first
employ the TST-MUSIC algorithm to obtain the estimate $\hat{\tau}_{k}^{(t)},\hat{\theta}_{k}^{(t)},\hat{\phi}_{k}^{(t)}$.
The signal model (\ref{eq:MUSIC_Signal}) can be reformulated as
\begin{equation}
\mathbf{y}^{(t)}=\mathbf{V}^{(t)}\widetilde{\boldsymbol{\alpha}}^{(t)}+\mathbf{n}^{(t)},
\end{equation}
where $\mathbf{y}^{(t)}\triangleq vec(\mathbf{Y}^{(t)})\in\mathbb{C}^{N_{r}P\times1}$,
$\mathbf{V}^{(t)}=\mathbf{A}(\theta,\phi)\odot\mathbf{G}^{(t)}(\boldsymbol{\tau}+\tau_{0}^{(t)})\in\mathbb{C}^{N_{r}P\times K}$,
$\mathbf{n}^{(t)}=vec(\mathbf{N}^{(t)})$, and $\widetilde{\boldsymbol{\alpha}}^{(t)}=[\widetilde{\alpha}_{1}^{(t)},...,\widetilde{\alpha}_{K}^{(t)}]^{T}$.
Then, we can obtain a LS solution of $\widetilde{\boldsymbol{\alpha}}^{(t)}$
as
\begin{equation}
\hat{\widetilde{\boldsymbol{\alpha}}}^{(t)}=(\mathbf{V}^{(t)H}\mathbf{V}^{(t)})^{-1}\mathbf{V}^{(t)H}\mathbf{y}^{(t)}.\label{eq:coarse_alpha}
\end{equation}

Finally, according to the inherent structure of $\widetilde{\boldsymbol{\alpha}}^{(t)}$,
the imperfection parameters can be estimated as\footnote{Note that we have absorb the term $e^{j\varphi_{1}^{(t)}}$ into $\varepsilon^{(t)}$
and estimate them as a whole, i.e., in fact, $\hat{\varepsilon}^{(t)}$
is the estimate of $e^{j(\varepsilon^{(t)}+\varphi_{1}^{(t)})}$ ,
and $\hat{\varphi}_{k}^{(t)}$ is the estimate of $\varphi_{k}^{(t)}-\varphi_{1}^{(t)}$.
This equivalent parameter estimation has no effect on the final channel
estimation performance.}
\begin{eqnarray}
\hat{\varepsilon}^{(t)} & = & \angle(\hat{\widetilde{\alpha}}_{1}^{(t)}/\hat{\widetilde{\alpha}}_{1}^{(1)}),\label{eq:coarse_fy}\\
\hat{\varphi}_{k}^{(t)} & = & \angle\frac{\hat{\widetilde{\alpha}}_{k}^{(t)}\hat{\widetilde{\alpha}}_{1}^{(1)}}{\hat{\widetilde{\alpha}}_{k}^{(1)}\hat{\widetilde{\alpha}}_{1}^{(t)}},\label{eq:coarse_Dop}\\
\hat{\tau}_{0}^{(t)} & = & \frac{\sum_{k=1}^{K^{(t)}}(\hat{\tau}_{k}^{(t)}-\hat{\tau}_{k}^{(1)})}{K^{(t)}}.\label{eq:Coarse_tau0}
\end{eqnarray}
\vspace{-0.1in}

\subsection{Refinement Estimation Phase}

In this stage, we perform a joint multi-timeslot channel parameter
estimation to achieve channel extrapolation after initialization.
The proposed algorithm executes AO iterations among four components
as follows.

\subsubsection{Step 1 (Multi-band observations splicing)}

\begin{figure}[t]
\begin{centering}
\textsf{\includegraphics[width=7cm]{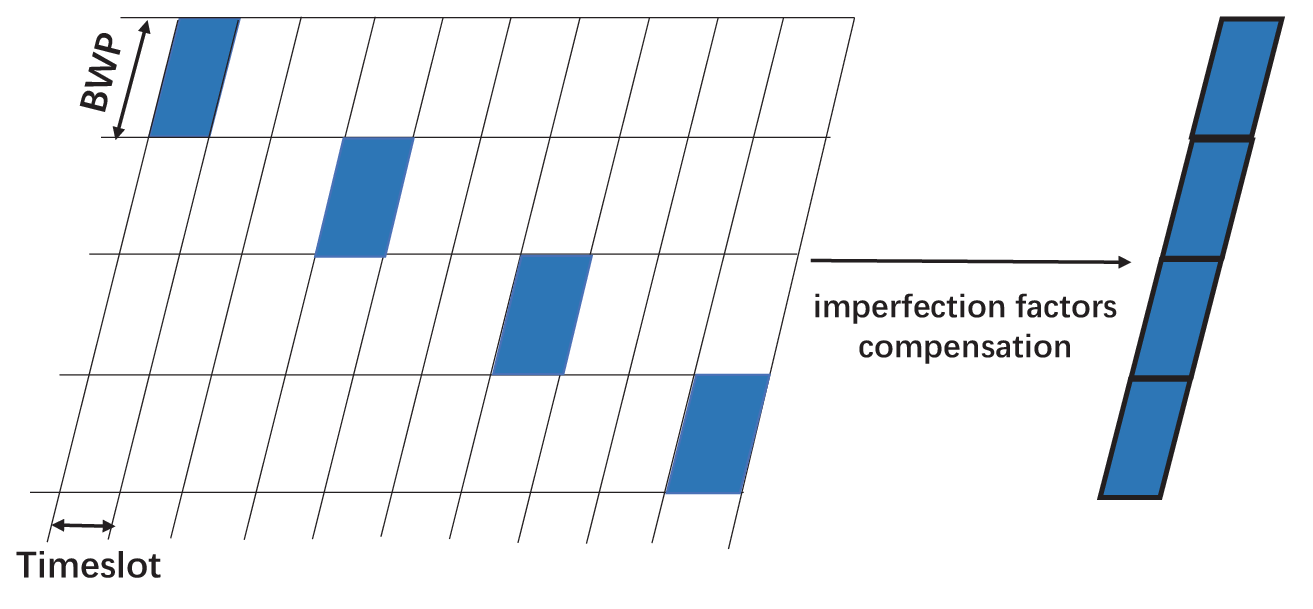}}
\par\end{centering}
\caption{\label{fig:step1}An illustration of step 1.}
\end{figure}

In this step, we compensate for the imperfection parameters and then
splice the observation samples obtained in different BWPs into a full-band
observation samples, as shown in Fig. \ref{fig:step1}.

For given estimated imperfection parameters $\hat{\varepsilon}^{(t)},\hat{\varphi}_{k}^{(t)},\hat{\tau}_{0}^{(t)}$,
we recover the ``clean'' observations as
\begin{equation}
\overline{\mathbf{Y}}^{(t)}=\textrm{diag}(e^{-j\hat{\varepsilon}^{(t)}})\cdot\textrm{diag}(\mathbf{W}^{(t)}\mathbf{a}(\hat{\tau}_{0}^{(t)})^{*})\cdot\mathbf{Y}^{(t)}.
\end{equation}
Then, we have a compensated full-band linear signal model given by
\begin{equation}
\overline{\mathbf{Y}}^{(T_{e})}=\overline{\mathbf{G}}^{(T_{e})}\left(\boldsymbol{\tau}\right)\overline{\mathbf{B}}\mathbf{A}\left(\theta,\phi\right)^{T}+\mathbf{N}^{(t)},\label{eq:compensated}
\end{equation}
where $\overline{\mathbf{Y}}^{(T_{e})}=[\overline{\mathbf{Y}}^{(1)};...;\overline{\mathbf{Y}}^{(T_{e})}]\in\mathbb{C}^{T_{e}P\times N_{r}}$,
$\overline{\mathbf{G}}^{(T_{e})}(\boldsymbol{\tau})=[\overline{\mathbf{G}}^{(1)}(\boldsymbol{\tau});...;\overline{\mathbf{G}}^{(T_{e})}(\boldsymbol{\tau})]\in\mathbb{C}^{T_{e}P\times K}$
with $\overline{\mathbf{G}}^{(t)}(\boldsymbol{\tau})=\mathbf{G}^{(t)}(\boldsymbol{\tau})\textrm{diag}(e^{j\hat{\varphi}_{1}^{(t)}},...,e^{j\hat{\varphi}_{K}^{(t)}})$,
and $\overline{\mathbf{B}}=\textrm{diag}(\alpha_{1},...,\alpha_{K})$.

\subsubsection{Step 2 (Joint delay-angle parameters estimation using TST-MUSIC-SIC)\label{subsec:Step-2}}

\begin{figure}[t]
\begin{centering}
\begin{minipage}[t]{0.45\textwidth}%
\begin{center}
\subfloat[\textcolor{blue}{\label{fig:MUSIC_pseudo_orth_a}}]{\begin{centering}
\includegraphics[width=8cm]{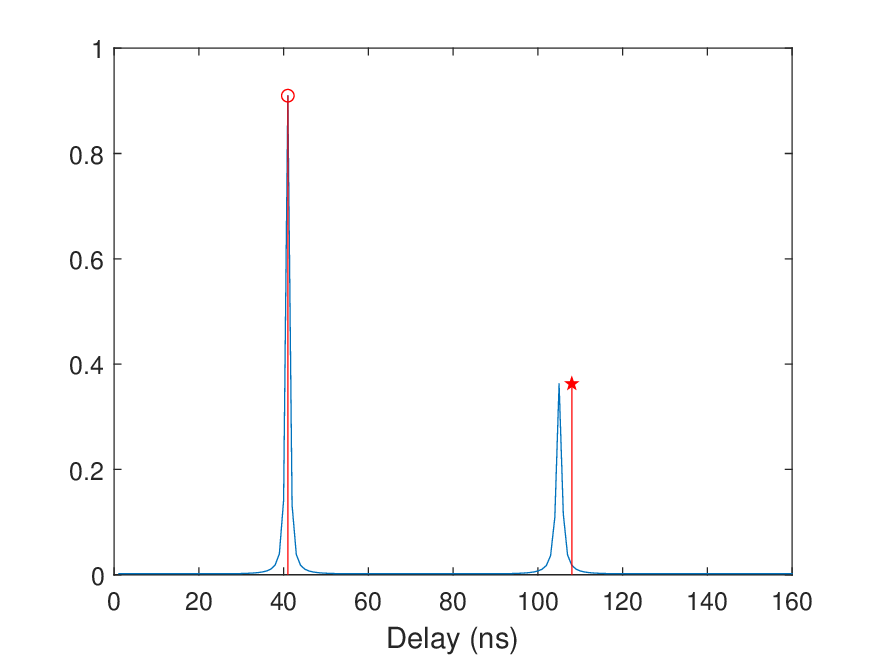}
\par\end{centering}
}
\par\end{center}%
\end{minipage}\hfill{}%
\begin{minipage}[t]{0.45\textwidth}%
\begin{center}
\subfloat[\textcolor{blue}{\label{fig:MUSIC_pseudo_orth_b}}]{\begin{centering}
\includegraphics[width=8cm]{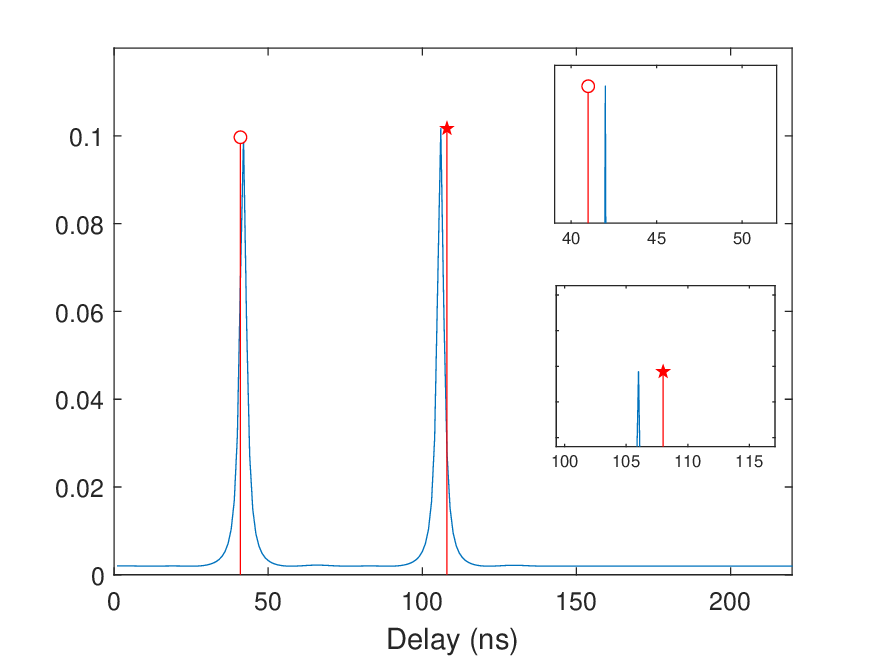}
\par\end{centering}
}
\par\end{center}%
\end{minipage}
\par\end{centering}
\medskip{}

\centering{}\caption{\label{fig:MUSIC_pseudo_orth}The pseudospectrum $\mathcal{\overline{P}}_{1}^{d}(\tau)$
in the cases: (a) Perfect imperfection parameters compensation; (b)
Imperfect imperfection parameters compensation.}
\end{figure}

Then, we apply the TST-MUSIC algorithm to the compensated observations
$\overline{\mathbf{Y}}^{(T_{e})}$ to estimate the delay and angle
parameters, but with a different T-MUSIC pseudospectrum estimation
method. Particularly, the primary orthogonality property in (\ref{eq:orthogonality_delay})
does not hold in the new full-band signal model (\ref{eq:compensated})
anymore due to the effect of $\hat{\varphi}_{k}^{(t)}$. Instead,
we have a new orthogonality property, $\overline{\mathbf{G}}^{(T_{e})\thinspace H}\mathbf{V}_{n}^{d}=\boldsymbol{O},$
i.e., $\overline{\mathbf{g}}_{k}(\tau_{k})^{H}\mathbf{V}_{n}^{d}=\mathbf{0}^{T},\forall k$,
where $\overline{\mathbf{g}}_{k}$ denotes the $k$-th column vector
of $\overline{\mathbf{G}}^{(T_{e})}$. Therefore, in contrast to the
primary T-MUSIC algorithm that $K$ delays are estimated based on
the same pseudospectrum, in our proposed R-TST-MUSIC algorithm, $K$
delays need to be estimated based on $K$ different pseudospectrums,
respectively. For instance, the delay $\tau_{k}$ can be estimated
at which the pseudospectrum $\mathcal{\overline{P}}_{k}^{d}(\tau)$
derived from the signal model (\ref{eq:compensated}) takes the maximum
value:
\begin{equation}
\mathcal{\overline{P}}_{k}^{d}(\tau)=\frac{1}{\overline{\mathbf{g}}_{k}(\tau)^{H}\left(\mathbf{I}-\mathbf{V}_{s}^{d}\mathbf{V}_{s}^{d\thinspace H}\right)\overline{\mathbf{g}}_{k}(\tau)},\forall k.\label{eq:Refine_spercturm_tau-1}
\end{equation}
However, for the delay estimate of a certain path based on (\ref{eq:Refine_spercturm_tau-1}),
the interference from other paths cannot be neglected, especially
in the delay estimation of a path with low-energy. In other words,
owing to the ``pseudo'' orthogonality, i.e., $\overline{\mathbf{g}}_{k}(\tau_{k^{\prime}})^{H}\left(\mathbf{I}-\mathbf{V}_{s}^{d}\mathbf{V}_{s}^{d\thinspace H}\right)\overline{\mathbf{g}}_{k}(\tau_{k^{\prime}})=\varepsilon,\forall k^{\prime}\neq k,$
where $\varepsilon$ has a small value, we may find a virtual delay
in the vicinity of $\tau_{k^{\prime}}$. To demonstrate this phenomenon
and highlight the challenges of multipaths delay estimation based
on (\ref{eq:Refine_spercturm_tau-1}), we present the curves of $\mathcal{\overline{P}}_{1}^{d}(\tau)$
in Fig. \ref{fig:MUSIC_pseudo_orth} for estimation of delay $\tau_{1}$
, where the red circle and the red star denote the true values of
$\tau_{1}$ and $\tau_{2}$, respectively. We set $K=2$, $h_{p}=4$,
the true delays $\tau_{1}=40$ ns with power $-8.8$ dB, $\tau_{2}=107$
ns with power 0 dB. As depicted in Fig. \ref{fig:MUSIC_pseudo_orth_a},
two peak are observed. The first peak appears in the true value of
 $\tau_{1}$, following the orthogonality property. However, another
peak with relatively low energy also emerges around the true value
of  $\tau_{2}$ due to the ``pseudo'' orthogonality, even if the
imperfection parameters are perfectly compensated for. However, as
shown in Fig. \ref{fig:MUSIC_pseudo_orth_b} when in the presence
of the imperfection parameters estimation errors, the first peak deviates
from the true value of  $\tau_{1}$. Moreover, there is a possibility
of identifying a wrong peak as the estimation result of $\tau_{1}$,
since the peak with the maximum energy may no longer be located around
$\tau_{1}$, but rather around $\tau_{2}$.

To solve this challenge, we adopt a successive interference cancellation
(SIC) method. We first estimate the delay of the path with the largest
energy ratio at which the following pseudospectrum takes the maximum
value
\begin{equation}
\mathcal{P}_{1}^{d}(\tau)=\frac{1}{\overline{\mathbf{g}}_{1}(\tau)^{H}\left(\mathbf{I}-\mathbf{V}_{s}^{d}\mathbf{V}_{s}^{d\thinspace H}\right)\overline{\mathbf{g}}_{1}(\tau)}.\label{eq:Refine_spercturm_tau}
\end{equation}
Without loss of generality, we have assumed that the estimated channel
gains of the multipaths have a descending order, i.e., $\left|\hat{\alpha}_{k}\right|^{2}\ge\left|\hat{\alpha}_{k+1}\right|^{2},\forall k\in\left\{ 1,...,K-1\right\} $.
Then, the contribution of the first estimated path is removed from
$\overline{\mathbf{Y}}^{(T_{e})}$, i.e.,
\begin{equation}
\overline{\mathbf{Y}}_{1}^{(T_{e})}=\mathbf{\overline{U}}_{1}^{d}\overline{\mathbf{Y}}^{(T_{e})},
\end{equation}
where $\mathbf{\overline{U}}_{1}^{d}=\mathbf{I}-\frac{1}{T_{e}P}\mathbf{\overline{g}}_{1}(\hat{\tau}_{1})\mathbf{\overline{g}}_{1}^{H}(\hat{\tau}_{1})$
denotes the temporal filtering matrix of the first path. Next, the
delay of the second strongest path $\tau_{2}$ is estimated based
on $\overline{\mathbf{Y}}_{1}^{(T_{e})}$, and so on. Using SIC method,
finally all delays $\tau_{1},...,\tau_{K}$ are estimated in turn.

Note that in the search of the peaks of the MUSIC pseudospectrum in
(\ref{eq:Refine_spercturm_tau}), the searching region can shrink
to the vicinity of the initial delay and angle estimation results
obtained from the initialization phase instead of the whole axis.
Thus, the computational complexity of this step can be significantly
reduced.

\subsubsection{Step 3 (Imperfection factors and channel coefficients estimation
based on ML)}

In this step, we employ ML method to estimate the imperfection parameters
and the channel coefficients. The optimization problem can be formulated
as
\begin{equation}
\begin{aligned} & \mathop{\arg\min}\limits _{\varphi_{k},\varepsilon^{(t)},\tau_{0}^{(t)},\alpha_{k}}\sum\limits _{t=1}^{T_{e}}\left\Vert \mathbf{Y}^{(t)}-\mathbf{G}^{(t)}\left(\hat{\boldsymbol{\tau}}+\tau_{0}^{(t)}\right)\mathbf{B}\left(\alpha_{k},\varphi_{k},\varepsilon^{(t)}\right)\right.\\
 & \thinspace\thinspace\thinspace\thinspace\thinspace\thinspace\thinspace\thinspace\thinspace\thinspace\thinspace\thinspace\thinspace\thinspace\thinspace\thinspace\thinspace\thinspace\thinspace\thinspace\thinspace\thinspace\thinspace\thinspace\thinspace\thinspace\thinspace\thinspace\thinspace\thinspace\thinspace\thinspace\thinspace\thinspace\thinspace\thinspace\thinspace\thinspace\thinspace\thinspace\thinspace\thinspace\thinspace\left.\times\mathbf{A}\left(\hat{\theta}_{k},\hat{\phi}_{k}\right)^{T}\right\Vert _{F}^{2}.
\end{aligned}
\label{eq:ML_Obj}
\end{equation}
Then, we use AO method to alternatively optimize the variables. Particularly,
we can obtain a closed-form solution for $\alpha_{k}$ and $\varepsilon^{(t)}$
as
\begin{eqnarray}
\hat{\boldsymbol{\alpha}} & = & ((\boldsymbol{\Psi}_{\alpha}^{(T_{e})})^{H}\boldsymbol{\Psi}_{\alpha}^{(T_{e})})^{-1}(\boldsymbol{\Psi}_{\alpha}^{(T_{e})})^{H}\mathbf{y}^{(T_{e})},\label{eq:refine_alpha}\\
\hat{\varepsilon}^{(t)} & = & \angle((\boldsymbol{\Psi}_{\varepsilon}^{(t)})^{H}\mathbf{y}^{(t)}),
\end{eqnarray}
where
\begin{align*}
\boldsymbol{\Psi}_{\alpha}^{(T_{e})} & =[\boldsymbol{\Psi}_{\alpha}^{(1)};...;\boldsymbol{\Psi}_{\alpha}^{(T_{e})}],\\
\!\!\boldsymbol{\Psi}_{\alpha}^{(t)} & =\mathbf{A}(\theta,\phi)\!\odot\!\mathbf{G}^{(t)}(\!\boldsymbol{\tau}\!+\!\tau_{0}^{(t)}\!)\!\cdot\!\textrm{diag}(\!e^{j\varphi_{k}^{(t)}}\!e^{j\varepsilon^{(t)}}\!\!\!\!,...,e^{j\varphi_{K}^{(t)}}\!e^{j\varepsilon^{(t)}}\!),\\
\boldsymbol{\Psi}_{\varepsilon}^{(t)} & =\mathbf{A}(\theta,\phi)\odot\mathbf{G}^{(t)}(\boldsymbol{\tau}+\tau_{0}^{(t)})\boldsymbol{\alpha}_{\varepsilon},\\
\boldsymbol{\alpha}_{\varepsilon} & =[\alpha_{1}e^{j\varphi_{1}^{(t)}},...,\alpha_{K}e^{j\varphi_{K}^{(t)}}]^{T}.
\end{align*}
Then, $\boldsymbol{\boldsymbol{\varphi}}$$\triangleq[\varphi_{1},$$...,$$\varphi_{K}]$
and $\tau_{0}^{(t)}$ can be estimated using gradient descent method
as
\begin{eqnarray}
\boldsymbol{\boldsymbol{\varphi}}^{(j+1)} & = & \boldsymbol{\boldsymbol{\varphi}}^{(j)}-\gamma_{\boldsymbol{\varphi}}\cdot\boldsymbol{\boldsymbol{\zeta}}_{\boldsymbol{\varphi}}^{(j)},\label{eq:updaterule_delta}\\
\tau_{0}^{(t)\thinspace\thinspace(j+1)} & = & \tau_{0}^{(t)\thinspace\thinspace(j)}-\gamma_{\tau_{0}^{(t)}}\cdot\zeta_{\tau_{0}^{(t)}}^{(j)},\label{eq:updaterule_tau}
\end{eqnarray}
where $\gamma_{\boldsymbol{\varphi}}$ and $\gamma_{\tau_{0}^{(t)}}$
are the step size determined by the Armijo rule \cite{Bertsekas_book95_NProgramming},
$\boldsymbol{\boldsymbol{\zeta}}_{\boldsymbol{\varphi}}^{(j)}$ and
$\zeta_{\tau_{0}^{(t)}}^{(j)}$ are the gradients of the objective
function in (\ref{eq:ML_Obj}) with respect to $\boldsymbol{\boldsymbol{\varphi}}$
and $\tau_{0}^{(t)}$, respectively.\vspace{-0.15in}

\subsection{Computational Complexity Analysis}

The main computational complexity of  R-TST-MUSIC depends on the eigendecomposition
of $\mathbf{R}^{d}$ and $\mathbf{R}^{s}$ based on $\overline{\mathbf{Y}}^{(T_{e})}$,
which are $\mathcal{O}\left(T_{e}^{3}P^{3}\right)$ and $\mathcal{O}\left(N_{r}^{3}\right)$,
respectively, the matrix multiplication and inverse operation in (\ref{eq:refine_alpha}),
which are $\mathcal{O}\left(T_{e}N_{r}PK^{2}\right)$ and $\mathcal{O}\left(K^{3}\right)$,
respectively. Besides, the computational complexity of the spatial
and temporal searches for the S-MUSIC and T-MUSIC pseudospectrum have
the orders of $\mathcal{O}\left(N_{r}^{2}g_{s}\right)$ and $\mathcal{O}\left(T_{e}^{2}P^{2}g_{t}\right)$,
respectively, where $g_{s}$ and $g_{t}$ are the numbers of searches
conducted along the angle axis and the delay axis.

As can be seen, the computational complexity in the channel estimation
stage has a cubic order of $T_{e}$, which is unacceptable when $T_{e}$
is large. Therefore, we further propose a channel tracking based extrapolation
scheme, which can exploit the time-correlation of the channel parameters
and avoid frequent multi-timeslot based channel estimation.\vspace{-0.15in}

\section{Channel Extrapolation in Channel Tracking Stage\label{sec:Tra}}

\begin{figure}[t]
\begin{centering}
\textsf{\includegraphics[width=8cm]{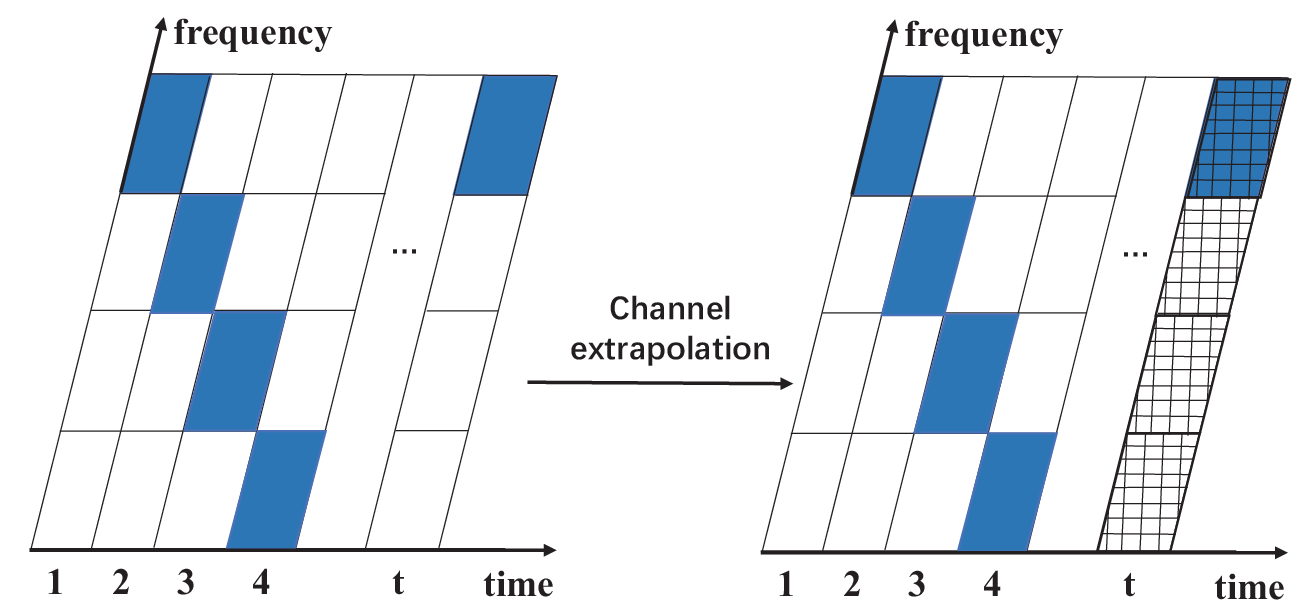}}
\par\end{centering}
\caption{\label{fig:extrapolation_tra}Channel extrapolation in channel tracking
stage.}
\end{figure}

In the channel tracking stage, we aim to achieve channel extrapolation
based on the received SRSs at the current time $t$ and the prior
information passed from time $t-1$, as shown in Fig. \ref{fig:extrapolation_tra}.
As compared to the one stage channel extrapolation scheme that performs
multi-timeslot channel extrapolation, our proposed two-stage scheme
is less time-consuming, especially for a long time channel estimation.
\vspace{-0.15in}

\subsection{Sparse Channel Representation}

We first describe sparse representation over delay and angular domain
for signal model (\ref{eq:recF}). One commonly used method is to
define a uniform delay grid $\mathcal{D}=\{\overline{d}_{1},\ldots,\overline{d}_{L}\}$
of $L$ ($L\gg K^{(t)},\forall t$) delay points over $[-\frac{1}{4}T_{d},T_{d}]$
($T_{d}$ denotes an upper bound for the maximum delay spread) and
two uniform angle grids $\mathcal{G}_{\theta}=\{\overline{\theta}_{1},\ldots,\overline{\theta}_{N_{x}}\}$
and $\mathcal{G}_{\phi}=\{\overline{\phi}_{1},\ldots,\overline{\phi}_{N_{y}}\}$
of $N_{x}$ and $N_{y}$ angle points over $\left[0,2\pi\right]$.
If all the true delay and angle values exactly lie in the discrete
sets $\mathcal{D}$ and $\mathcal{G}_{\theta},\mathcal{\mathcal{G}_{\phi}}$,
we can reformulate signal model (\ref{eq:recF}) as
\begin{equation}
\mathbf{Y}^{(t)}=\textrm{diag}\left(\mathbf{x}\right)\mathbf{W}^{(t)}\boldsymbol{\Lambda}^{(t)}\mathbf{S}^{(t)}\mathbf{F}_{d}(\mathbf{H}_{d}^{(t)}\ast\mathbf{D}^{(t)})\mathbf{A}_{R}^{T}+\mathbf{N}^{(t)},\label{eq:D-A Channel}
\end{equation}
where $\boldsymbol{\Lambda}^{(t)}=\mathrm{diag}(e^{j\varepsilon^{(t)}},\ldots,e^{j\varepsilon^{(t)}}),$
\[
\begin{aligned}\mathbf{S}^{(t)} & =\mathrm{diag}(1,\ldots,e^{-j2\pi nf_{s}\tau_{0}^{(t)}},\ldots,e^{-j2\pi(N-1)f_{s}\tau_{0}^{(t)}}),\\
\mathbf{F}_{d} & =[\mathbf{a}(\overline{d}_{1}),\mathbf{a}(\overline{d}_{2}),\cdots,\mathbf{a}(\overline{d}_{L})],\\
\mathbf{A}_{R} & =[\boldsymbol{a}_{R}(\overline{\theta}_{1},\overline{\phi}_{1}),\cdots,\boldsymbol{a}_{R}(\overline{\theta}_{n_{x}},\overline{\phi}_{n_{y}}),\cdots,\boldsymbol{a}_{R}(\overline{\theta}_{N_{x}},\overline{\phi}_{N_{y}})].
\end{aligned}
\]
The matrix $\mathbf{F}_{d}\in\mathbb{C}^{N\times L}$ and $\boldsymbol{A}_{R}\in\mathbb{C}^{N_{r}\times N_{r}}$
denote the dictionary matrix consisting of linear steering vectors
evaluated on delay grids $\mathcal{D}$ and angle grids $\mathcal{G}_{\theta},\mathcal{G}_{\phi}$,
respectively. The matrix $\boldsymbol{\Lambda}^{(t)}\in\mathbb{C}^{N\times N}$
and $\mathbf{S}^{(t)}\in\mathbb{C}^{N\times N}$ denote the diagonal
matrix associated with imperfection factors $\varepsilon^{(t)}$ and
$\tau_{0}^{(t)}$, $\mathbf{D}^{(t)}\in\mathbb{C}^{L\times N_{r}}$
denotes the matrix corresponding to $\varphi_{k}^{(t)}$, and $\mathbf{H}_{d}^{(t)}$
denotes the sparse delay-angular domain (DAD) channel matrix whose
non-zero elements correspond to the true paths. 

However,  the delay and angle resolution of the algorithm designed
from the on-grid signal model (\ref{eq:D-A Channel}) are limited
to the grid spacing, which results in a significant performance loss
for channel extrapolation. To handle this issue,  we introduce a delay
off-grid vector $\Delta\boldsymbol{\tau}^{(t)}=[\Delta\tau_{1}^{(t)},\cdots,\Delta\tau_{L}^{(t)}]$
satisfying $\Delta\tau_{l_{k}}^{(t)}=\tau_{k}^{(t)}-\overline{d}_{l_{k}},k=1,\cdots,K^{(t)}$,
and $\Delta\tau_{l}^{(t)}=0,\forall l\notin\left\{ l_{1},\cdots,l_{K^{(t)}}\right\} $,
where $l_{k}\triangleq\underset{l}{\mathrm{argmin}}\left|\tau_{k}^{(t)}-\overline{d}_{l}\right|$
denotes the index of grid which is nearest to $\tau_{k}^{(t)}$. Furthermore,
we formulate a probability model of the off-grid vector in order to
capture its time-correlation property, as detailed later. We also
introduce angle off-grid vectors $\Delta\boldsymbol{\theta}^{(t)}=[\Delta\theta_{1}^{(t)},\cdots,\Delta\theta_{N_{x}}^{(t)}]$
and $\Delta\boldsymbol{\phi}^{(t)}=[\Delta\phi_{1}^{(t)},\cdots,\Delta\phi_{N_{y}}^{(t)}]$
for $\mathbf{A}_{R}$ in a manner similar to the delay off-grid vector.
For convenience, we  denote all the off-grid parameter vectors as
$\boldsymbol{\varDelta}^{(t)}\triangleq[\Delta\boldsymbol{\tau}^{(t)},\Delta\boldsymbol{\theta}^{(t)},\Delta\boldsymbol{\phi}^{(t)}]$.
Finally, the dictionary matrix $\mathbf{F}_{d}$ and $\mathbf{A}_{R}$
can be rewritten as
\begin{align*}
 & \mathbf{F}_{d}\left(\Delta\boldsymbol{\tau}^{(t)}\right)=\left[\mathbf{a}\left(\overline{d}_{1}+\Delta\tau_{1}^{(t)}\right),\cdots,\mathbf{a}\left(\overline{d}_{L}+\Delta\tau_{L}^{(t)}\right)\right],\\
 & \mathbf{A}_{R}\left(\Delta\boldsymbol{\theta}^{(t)},\Delta\boldsymbol{\phi}^{(t)}\right)=\left[\boldsymbol{a}_{R}(\overline{\theta}_{1}+\Delta\theta_{1}^{(t)},\overline{\phi}_{1}+\Delta\phi_{1}^{(t)}),\cdots,\right.\\
 & \left.\thinspace\thinspace\thinspace\thinspace\thinspace\thinspace\thinspace\thinspace\thinspace\thinspace\thinspace\thinspace\thinspace\thinspace\thinspace\thinspace\thinspace\thinspace\thinspace\thinspace\thinspace\thinspace\thinspace\thinspace\thinspace\thinspace\thinspace\thinspace\thinspace\thinspace\thinspace\thinspace\thinspace\thinspace\thinspace\thinspace\thinspace\thinspace\boldsymbol{a}_{R}(\overline{\theta}_{N_{x}}+\Delta\theta_{N_{x}}^{(t)},\overline{\phi}_{N_{y}}+\Delta\phi_{N_{y}}^{(t)})\right].
\end{align*}
\vspace{-0.15in}

\subsection{Probability Model}

In this subsection, we propose a Markov channel model to capture the
dynamic sparsity of the DAD channel vector $\boldsymbol{h}_{d}^{(t)}\triangleq vec(\mathbf{H}_{d}^{(t)})$,
and time-correlation of the off-grid vectors $\boldsymbol{\varDelta}^{(t)}$
and Doppler frequency offset $\boldsymbol{f}_{D}^{(t)}\in\mathbb{R}^{LN_{r}}$
with its element $f_{D,m}^{(t)}$ denoting the Doppler frequency offset
associated with DAD channel coefficient $h_{d,m}^{(t)}$. For convenience,
we denote a time series of DAD channels $\{\boldsymbol{h}_{d}^{(1)},...,\boldsymbol{h}_{d}^{(t)}\}$
in channel tracking stage as $\boldsymbol{h}_{d}^{(T)}$ (same for
$\boldsymbol{\vartheta}^{(T)}$, $\boldsymbol{s}^{(T)}$, $\boldsymbol{f}_{D}^{(T)}$,
$\boldsymbol{\varDelta}^{(T)}$). 

\subsubsection{Probability Model for DAD Channel $\boldsymbol{h}_{d}^{(T)}$}

To capture the temporal correlation and promote sparsity of the DAD
channel $\boldsymbol{h}_{d}^{(T)}$, we employ a widely used Bernoulli-Gaussian
(BG) probability model, which can be written as \cite{BG_3,BG_1,BG_2}
\begin{eqnarray*}
p\left(\boldsymbol{h}_{d}^{(T)}|\boldsymbol{\vartheta}^{(T)},\boldsymbol{s}^{(T)}\right) & = & \prod_{t=1}^{T}\prod_{m=1}^{LN_{r}}p\left(h_{d,m}^{(t)}|s_{m}^{(t)},\vartheta_{m}^{(t)}\right)\\
 & = & \prod_{t=1}^{T}\prod_{m=1}^{LN_{r}}\delta\left(h_{d,m}^{(t)}-s_{m}^{(t)}\vartheta_{m}^{(t)}\right),
\end{eqnarray*}
where $s_{m}^{(t)}\in\{0,1\}$ describes the birth-death process of
the multipath and $\vartheta_{m}^{(t)}$ describes the smooth evolution
of the amplitudes of the non-zero channel coefficients. Then, the
sparse Markov channel prior distribution is given by
\begin{alignat}{1}
 & p\!\left(\!\boldsymbol{h}_{d}^{(T)}\!,\boldsymbol{\vartheta}^{(T)}\!,\boldsymbol{s}^{(T)}\!,\boldsymbol{f}_{D}^{(T)}\!,\boldsymbol{\varDelta}^{(T)}\!\right)\!=\!p\left(\!\boldsymbol{f}_{D}^{(T)}\!\right)p\left(\!\Delta\mathbf{\boldsymbol{\tau}}^{(T)}\!\right)p\left(\!\Delta\boldsymbol{\phi}^{(T)}\!\right)\nonumber \\
 & \thinspace\thinspace\thinspace\thinspace\thinspace\thinspace\times p\left(\Delta\boldsymbol{\theta}^{(T)}\right)p\left(\boldsymbol{s}^{(T)}\right)p\left(\boldsymbol{\vartheta}^{(T)}\right)p\left(\boldsymbol{h}_{d}^{(T)}|\boldsymbol{\vartheta}^{(T)},\boldsymbol{s}^{(T)}\right).\label{eq:three layer}
\end{alignat}

\subsubsection{Probability Model for Channel Support $\boldsymbol{s}^{(T)}$}

Due to the slowly varying propagation environment, the channel supports
vary slowly over time. We use a Markov chain to model the temporal
correlation of the variables $\boldsymbol{s}^{(T)}$ as
\begin{alignat}{1}
p\left(\boldsymbol{s}^{\left(T\right)}\right)=\prod_{m=1}^{LN_{r}}p\left(s_{m}^{(1)}\right) & \prod_{t=2}^{T}p\left(s_{m}^{(t)}|s_{m}^{(t-1)}\right),\label{eq:Markov support}
\end{alignat}

\noindent where the transition probability $p\left(s_{m}^{(t)}=1|s_{m}^{(t-1)}=0\right)=\rho_{01}$,
and $p\left(s_{m}^{(t)}=0|s_{m}^{(t-1)}=1\right)=\rho_{10}$. The
Markov parameters $\left\{ \rho_{10},\rho_{01}\right\} $ characterize
the degree of temporal correlation of the channel support, e.g., smaller
$\rho_{10}$ or $\rho_{01}$ leads to highly correlated supports across
time, which means the propagation environment between the user and
BS varies slowly. 

\subsubsection{Probability Model for Hidden Variable $\boldsymbol{\vartheta}^{(T)}$}

The amplitude of path gains evolves smoothly over time and thus has
a temporal structure to be exploited. We use the Gauss-Markov processes
to model the temporal evolution of $\boldsymbol{\vartheta}^{(t)}$
as \cite{BG_3,BG_1}
\[
\vartheta_{m}^{(t)}=\left(1-\beta_{\vartheta}\right)\left(\vartheta_{m}^{(t-1)}-\mu_{\vartheta}\right)+\beta_{\vartheta}\omega_{m}^{(t)}+\mu_{\vartheta},m=1,...,LN_{r},
\]
where $\beta_{\vartheta}\in\left[0,1\right]$ controls the temporal
correlation, $\mu_{\vartheta}$ is the steady-state mean of the process,
and $\omega_{m}^{(t)}\sim\mathcal{CN}(0,\gamma_{\vartheta})$ is an
i.i.d. circular white Gaussian perturbation. Then, the joint distribution
$p(\boldsymbol{\vartheta}^{(T)})$ can be formulated as
\begin{equation}
p(\boldsymbol{\vartheta}^{(T)})=\prod_{m=1}^{LN_{r}}p\left(\vartheta_{m}^{(1)}\right)\prod_{t=2}^{T}p\left(\vartheta_{m}^{(t)}|\vartheta_{m}^{(t-1)}\right),\label{eq:hidden_value}
\end{equation}
where $p(\!\vartheta_{m}^{(t)}|\vartheta_{m}^{(t-1)}\!)\!\sim\mathcal{CN}\!(\!\vartheta_{m}^{(t)}\text{;(1-\ensuremath{\beta_{\vartheta}})}\vartheta_{m}^{(t-1)}\!\!+\!\beta_{\vartheta}\mu_{\vartheta},\beta_{\vartheta}^{2}\gamma_{\vartheta}\!)$.

\subsubsection{Probability Model for Doppler Frequency Offset $\boldsymbol{f}_{D}^{(T)}$}

The probability model is given by\vspace{-0.1cm}
\begin{equation}
p(\boldsymbol{f}_{D}^{(T)})=p(\boldsymbol{f}_{D}^{(1)})\prod_{t=2}^{T}p(\boldsymbol{f}_{D}^{(t)}|\boldsymbol{f}_{D}^{(t-1)}),\label{eq:imperfections}
\end{equation}
where $p(\boldsymbol{f}_{D}^{(t)}|\boldsymbol{f}_{D}^{(t-1)})$ is
the transition probability of  $\boldsymbol{f}_{D}^{(t)}$. We assume
that the user\textquoteright s acceleration is small and thus  $\boldsymbol{f}_{D}^{(t)}$
has high correlation over time. As such, we can use Gauss-Markov processes
to capture the time correlation of $\boldsymbol{f}_{D}^{(t)}$ as
\[
f_{D,m}^{(t)}\!=\!(1\!-\beta_{D})\!(f_{D,m}^{(t-1)}\!-\!\mu_{D})+\beta_{D}\omega_{m}^{(t)}+\mu_{D},m=1,...,LN_{r},
\]
with transition probability\vspace{-0.1cm}
\begin{align*}
 & p\left(\boldsymbol{f}_{D}^{\left(t\right)}|\boldsymbol{f}_{D}^{\left(t-1\right)}\right)\\
= & \prod_{m=1}^{LN_{r}}\mathcal{CN}\left(f_{D,m}^{(t)}\text{;\ensuremath{\left(1-\beta_{D}\right)}}f_{D,m}^{(t-1)}+\beta_{D}\mu_{D},\beta_{D}^{2}\gamma_{D}\right),
\end{align*}
where $\beta_{D}$, $\mu_{D}$, and $\gamma_{D}$ have similar definitions
with $\beta_{\vartheta}$, $\mu_{\vartheta}$, and $\gamma_{\vartheta}$
mentioned above.

\subsubsection{Probability Model for off-grid vectors $\boldsymbol{\varDelta}^{(T)}$}

Dynamic channel parameter estimation algorithms that based on the
off-grid model, typically utilize the EM method to estimate off-grid
vectors without using any prior information \cite{lian2019exploiting,BG_1}.
However, in the scenario of channel extrapolation that has a high
requirement on the channel parameter estimation accuracy, we are supposed
to fully exploit the time-correlation of the channel parameter to
improve the channel extrapolation performance by passing the high-resolution
prior information of the off-grid vectors from the previous timeslot.
Thus, we further formulate a probability model for the off-grid vectors
$\boldsymbol{\varDelta}^{(T)}$ as the delay and angle parameters
vary smoothly over time, which can be formulated as \cite{Kalman_1}
\begin{eqnarray*}
\Delta\boldsymbol{\tau}^{(t)} & = & \Delta\mathbf{\boldsymbol{\tau}}^{(t-1)}+\boldsymbol{u}_{\tau}^{(t)},\\
\Delta\boldsymbol{\theta}^{(t)} & = & \Delta\boldsymbol{\theta}^{(t-1)}+\boldsymbol{u}_{\theta}^{(t)},\\
\Delta\boldsymbol{\phi}^{(t)} & = & \Delta\boldsymbol{\phi}^{(t-1)}+\boldsymbol{u}_{\phi}^{(t)},
\end{eqnarray*}
where $\boldsymbol{u}_{\tau}^{(t)}$, $\boldsymbol{u}_{\theta}^{(t)}$,
and $\boldsymbol{u}_{\phi}^{(t)}$ $\sim\mathcal{CN}(0,\gamma_{u}\mathbf{I})$
denote the Gaussian noise. Then, the joint distribution can be formulated
as\vspace{-0.1cm}
\begin{equation}
\begin{aligned} & p\left(\Delta\boldsymbol{\tau}^{\left(T\right)}\right)=\prod_{m=1}^{L}p(\Delta\tau_{m}^{(1)})\prod_{t=2}^{T}p\left(\Delta\tau_{m}^{(t)}|\Delta\tau_{m}^{(t-1)}\right),\\
 & p\left(\Delta\boldsymbol{\theta}^{\left(T\right)}\right)=\prod_{m=1}^{N_{x}}p(\Delta\theta_{m}^{(1)})\prod_{t=2}^{T}p\left(\Delta\theta_{m}^{(t)}|\Delta\theta_{m}^{(t-1)}\right),\\
 & p\left(\Delta\boldsymbol{\phi}^{\left(T\right)}\right)=\prod_{m=1}^{N_{y}}p(\Delta\phi_{m}^{(1)})\prod_{t=2}^{T}p\left(\Delta\phi_{m}^{(t)}|\Delta\phi_{m}^{(t-1)}\right),
\end{aligned}
\label{eq:offgrid-1}
\end{equation}
where $p(\Delta\tau_{m}^{(t)}|\Delta\tau_{m}^{(t-1)})\sim\mathcal{N}(\Delta\tau_{m}^{(t)}\text{;}\Delta\tau_{m}^{(t-1)},\gamma_{u})$,
$p($$\Delta$$\theta_{m}^{(t)}|$$\Delta$$\theta_{m}^{(t-1)})$
and $p(\Delta\phi_{m}^{(t)}|\Delta\phi_{m}^{(t-1)})$ have similar
distributions.\vspace{-0.15in}

\subsection{Outline for Channel Tracking Algorithm}

\begin{figure*}[t]
\begin{centering}
\textsf{\includegraphics[width=15cm]{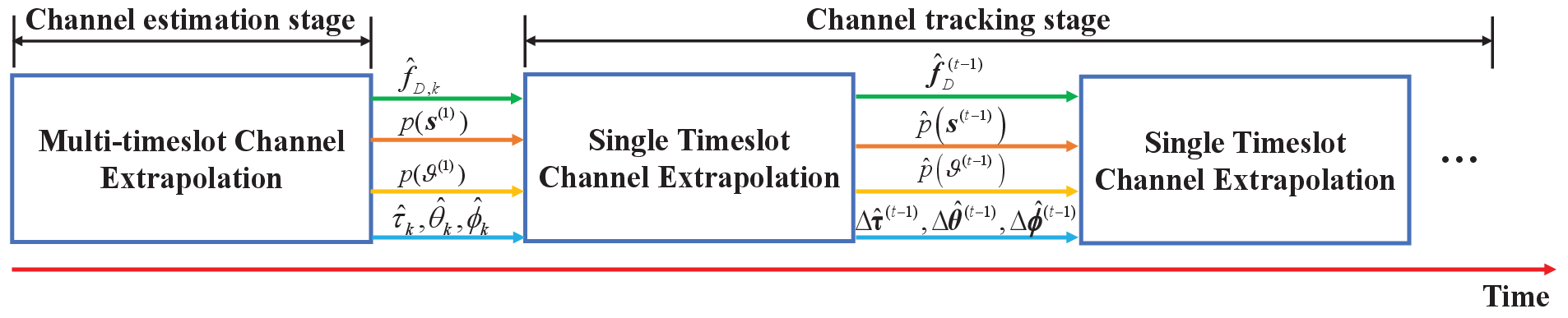}}
\par\end{centering}
\caption{\label{fig:flow}The flow chart of the proposed two-stage channel
extrapolation scheme.}
\end{figure*}

A flow chart depicting the prior information flow of the proposed
two-stage channel extrapolation scheme is shown in Fig. \ref{fig:flow}.
We map the paths estimated from the channel estimation stage into
the delay-angular domain with $i_{k}$ denoting the index corresponding
to the $k$-th estimated path. Then, we initialize the channel tracking
stage as\vspace{-0.1cm}
\begin{align*}
p(s_{m}^{(1)}) & =\begin{cases}
(\rho_{10})^{1-s_{m}^{(1)}}\!(1\!-\!\rho_{10})^{s_{m}^{(1)}} & ,m=i_{k}\\
(\rho_{01})^{s_{m}^{(1)}}\!(1\!-\!\rho_{01})^{1-s_{m}^{(1)}} & ,m\neq i_{k}
\end{cases},\\
p(\vartheta_{m}^{(1)}) & =\begin{cases}
\mathcal{CN}(\vartheta_{m}^{(1)};(1\!-\beta_{\vartheta})\hat{\alpha}_{k}\!+\!\beta_{\vartheta}\mu_{\vartheta},\beta_{\vartheta}^{2}\gamma_{\vartheta}) & ,m=i_{k}\\
\mathcal{CN}(\vartheta_{m}^{(1)};\beta_{\vartheta}\mu_{\vartheta},\beta_{\vartheta}^{2}\gamma_{\vartheta}) & ,m\neq i_{k}
\end{cases},
\end{align*}
and $\hat{f}_{D,i_{k}}^{(1)}=\hat{f}_{D,k},\forall k,$ the delay
off-grid vector $\Delta\hat{\tau}_{l_{k}}^{(1)}=\hat{\tau}_{k}-\overline{d}_{l_{k}},\forall k$
(angle off-grid vectors are initialized in a similar manner), where
$p(\boldsymbol{s}^{(1)})$ and $p(\boldsymbol{\vartheta}^{(1)})$
denote the initial distribution of the channel support and hidden
variable. Then, based on the proposed Markov probability model, the
prior information $p(\boldsymbol{s}^{(1)})$, $p(\boldsymbol{\vartheta}^{(1)})$,
and $\hat{f}_{D,k},\hat{\tau}_{k},\hat{\theta}_{k},\hat{\phi}_{k}$
can be exploited in the E-Step and M-Step of the proposed channel
tracking algorithm at time $t$, respectively.

For the prior information passing inside the channel tracking stage,
the prior information for time $t$ is the estimated posterior distribution
$\hat{p}(\boldsymbol{s}^{(t-1)})$, $\hat{p}(\boldsymbol{\vartheta}^{(t-1)})$
as in (\ref{eq:ps_pvm}), and point estimation results $\hat{\boldsymbol{f}}_{D}^{(t-1)},$$\Delta\hat{\boldsymbol{\tau}}^{\left(t-1\right)},$$\Delta\hat{\boldsymbol{\theta}}^{(t-1)},$$\Delta\hat{\boldsymbol{\phi}}^{(t-1)}$
passed from time $t-1$, which are also captured by the proposed Markov
probability model  and then exploited in the E-Step and M-Step of
the proposed channel tracking algorithm at time $t$, respectively.

The received signal model (\ref{eq:D-A Channel}) can be rewritten
as a standard CS model\vspace{-0.15cm}
\begin{align}
\mathbf{y}^{(t)} & =\boldsymbol{\Phi}^{(t)}\boldsymbol{h}_{d}^{(t)}+\mathbf{n}^{(t)},\label{eq:CS model}
\end{align}
where $\boldsymbol{\Phi}^{(t)}=[\boldsymbol{A}_{R}\otimes(\textrm{diag}(\mathbf{x})\mathbf{W}^{(t)}\boldsymbol{\Lambda}^{(t)}\mathbf{S}^{(t)}\mathbf{F}_{d})]\mathbf{D}^{(t)}\in\mathbb{C}^{N_{r}P\times N_{r}L}$.
At the $t$-th SRS symbol, we aim to track the time-varying DAD channel
$\boldsymbol{h}_{d}^{(t)}$, the off-grid vector $\boldsymbol{\varDelta}^{(t)}$,
and the imperfection parameter $\boldsymbol{\xi}^{(t)}\triangleq\{\boldsymbol{f}_{D}^{(t)},\varepsilon^{(t)},\tau_{0}^{(t)}\}$
based on the observations $\mathbf{y}^{(T)}$. In particular, given
the imperfection parameters $\boldsymbol{\xi}^{(t)}$ and the off-grid
vector $\boldsymbol{\varDelta}^{(t)}$, we are interested in computing
the MMSE estimates of $\boldsymbol{h}_{d}^{(t)}$, i.e., $\hat{h}_{d,m}^{(t)}=\mathrm{\mathbf{E}}\left[h_{d,m}^{(t)}|\mathbf{y}^{(T)};\boldsymbol{\xi}^{(t)},\boldsymbol{\varDelta}^{(t)}\right],m=1,...,LN_{r}$,
where the expectation is over the marginal posterior\vspace{-0.15cm}
\begin{flalign}
 & p\left(h_{d,m}^{(t)}|\mathbf{y}^{(T)};\boldsymbol{\xi}^{(t)},\boldsymbol{\varDelta}^{(t)}\right)\propto\nonumber \\
 & \thinspace\thinspace\thinspace\thinspace\sum_{\boldsymbol{s}^{(T)}}\int_{-h_{d,m}^{(t)},\boldsymbol{\vartheta}^{(T)}}p\left(\mathbf{y}^{(T)},\boldsymbol{v}^{(T)};\boldsymbol{\xi}^{(t)},\boldsymbol{\varDelta}^{(t)}\right),\label{eq:posterior}
\end{flalign}
where $\boldsymbol{v}^{(T)}$ denotes collections $\{\boldsymbol{v}^{(t)}\}_{t=1}^{T}$
with $\boldsymbol{v}^{(t)}\triangleq[\boldsymbol{h}_{d}^{(t)},\boldsymbol{\vartheta}^{(t)},\boldsymbol{s}^{(t)}]$,
$\int_{-h_{d,m}^{(t)}}$ denotes the vector collections integration
over $\boldsymbol{h}_{d}^{(T)}$ excluding the element $h_{d,m}^{(t)}$.
It is difficult to calculate the exact posterior in (\ref{eq:posterior})
because the corresponding factor graph has loops. Consequently, we
propose an efficient tracking algorithm combining the sum-product
message-passing (SPMP) algorithm \cite{SPMP} and the turbo framework
\cite{Turbo_CS} to calculate an approximate marginal posterior of
$p(h_{d,m}^{(t)}|\mathbf{y}^{(T)};\boldsymbol{\xi}^{(t)},\boldsymbol{\varDelta}^{(t)})$.

Besides, the imperfection parameters $\boldsymbol{\xi}^{(t)}$ and
the off-grid parameters $\boldsymbol{\varDelta}^{(t)}$ at time $t$
can be obtained by the maximum a posterior (MAP) estimator as follows:\vspace{-0.15cm}
\begin{alignat}{1}
\hat{\boldsymbol{\Xi}}^{(t)} & =\mathrm{arg}\max_{\boldsymbol{\Xi}^{(t)}}\ln p\left(\boldsymbol{\Xi}^{(t)}|\mathbf{y}^{(T)}\right)\nonumber \\
 & \propto\mathrm{arg}\max_{\boldsymbol{\Xi}^{(t)}}\ln\int p\left(\boldsymbol{h}_{d}^{(T)},\mathbf{y}^{(T)},\boldsymbol{\Xi}^{(t)}\right)d\boldsymbol{h}_{d}^{(T)},\label{eq:maxlikelyhood}
\end{alignat}
where $\boldsymbol{\Xi}^{(t)}\triangleq\{\boldsymbol{\xi}^{(t)},\boldsymbol{\varDelta}^{(t)}\}$.
It is a high-dimensional non-convex objective function and we cannot
obtain a closed-form expression due to the multi-dimensional integration
over $\boldsymbol{h}_{d}^{(T)}$. To handle this issue, we adopt majorization-minimization
(MM) method to construct a surrogate function and then use AO method
to find a stationary point of (\ref{eq:maxlikelyhood}). Inspired
by the EM method \cite{2003EM}, the channel tracking algorithm performs
iterations between following two steps until convergence at each time
$t$.
\begin{itemize}
\item \textbf{DAD Channel Estimation  (E-step):}\textsl{ }Given $\boldsymbol{\Xi}^{(t)}$,
 calculate the approximate posterior $\hat{p}\left(h_{d,m}^{(t)}|\mathbf{y}^{(T)};\boldsymbol{\Xi}^{(t)}\right)$
via the dynamic Turbo framework, as elaborated in Subsection \ref{subsec:E-STEP}.
\item \textbf{Off-grid and Imperfection Parameters Estimation  (M-step):}
Given the estimate of $\boldsymbol{\Xi}^{(t-1)}$ from the previous
time $t-1$, $\hat{p}(\boldsymbol{h}_{d}^{(t)}|\mathbf{y}^{(T)};\boldsymbol{\Xi}^{(t)})$
from the E-Step, construct surrogate functions for the MAP objective
function in (\ref{eq:maxlikelyhood}), then maximize the surrogate
function with respect to $\boldsymbol{\Xi}^{(t)}$, as elaborated
in Subsection \ref{sec:M-step}.\vspace{-0.1in}
\end{itemize}

\subsection{E-Step\label{subsec:E-STEP}}

\begin{figure}[t]
\begin{centering}
\textsf{\includegraphics[width=8cm]{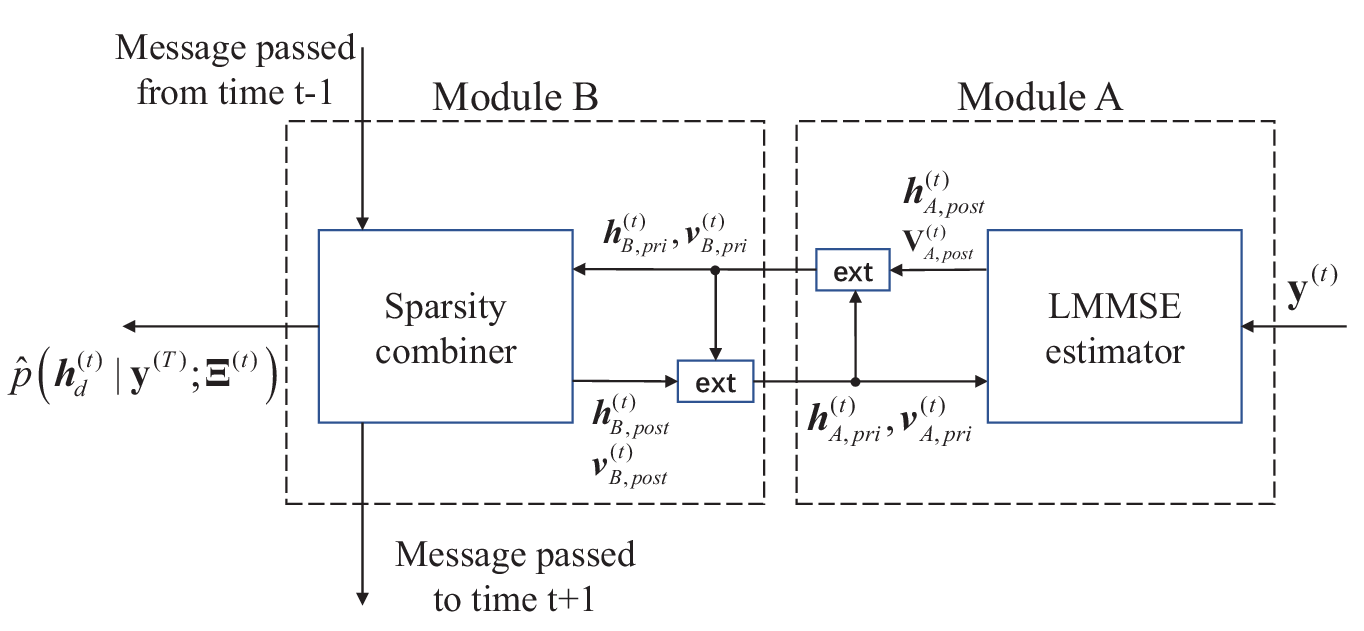}}
\par\end{centering}
\caption{\label{fig:turbo}An illustration of the Turbo-CS framework.}
\end{figure}

At the $t$-th SRS symbol, the E-Step contains two modules based on
the Turbo-CS framework as shown in Fig. \ref{fig:turbo}: Module A
is a linear minimum mean square error (LMMSE) estimator based on the
current observation $\mathbf{y}^{(t)}$ and messages from Module B,
while Module B is a sparsity combiner performing MMSE estimation that
combines the channel prior passed from the previous timeslot and the
messages from Module A. The two modules are executed iteratively until
convergence. 

\subsubsection{Module A}

In Module A, the DAD channel vector $\boldsymbol{h}_{d}^{(t)}$ is
estimated based on the current observation $\mathbf{y}^{(t)}$ and
a prior distribution $\mathcal{CN}(\boldsymbol{h}_{d}^{(t)};\boldsymbol{h}_{A,pri}^{(t)},\textrm{diag}(\boldsymbol{v}_{A,pri}^{(t)}))$,
where $\boldsymbol{h}_{A,pri}^{(t)}$ and $\boldsymbol{v}_{A,pri}^{(t)}$
are the extrinsic message output from Module B. Then, the LMMSE estimate
of $\boldsymbol{h}_{d}^{(t)}$ still follows a complex Gaussian distribution
with mean and variance given by\vspace{-0.1in}
\begin{equation}
\begin{aligned} & \boldsymbol{V}_{A,post}^{\left(t\right)}=\left((\boldsymbol{\Phi}^{(t)\thinspace H}\boldsymbol{\Phi}^{(t)})/(\sigma_{e}^{(t)})^{2}+\textrm{diag}(1/\boldsymbol{v}_{A,pri}^{(t)})\right)^{-1},\\
 & \boldsymbol{h}_{A,post}^{\left(t\right)}=\boldsymbol{V}_{A,post}^{\left(t\right)}\left(\frac{\boldsymbol{h}_{A,pri}^{(t)}}{\boldsymbol{v}_{A,pri}^{(t)}}+\frac{\boldsymbol{\Phi}^{(t)H}\mathbf{y}^{(t)}}{(\sigma_{e}^{(t)})^{2}}\right).
\end{aligned}
\label{eq:Va_post}
\end{equation}
Then, the extrinsic message passed to Module B can be calculated as
\cite{Turbo_CS}\vspace{-0.06in}
\begin{equation}
\begin{aligned} & \boldsymbol{h}_{B,pri}^{\left(t\right)}=\boldsymbol{v}_{B,pri}^{\left(t\right)}\left(\boldsymbol{h}_{A,post}^{\left(t\right)}/\boldsymbol{v}_{A,post}^{\left(t\right)}-\boldsymbol{h}_{A,pri}^{\left(t\right)}/\boldsymbol{v}_{A,pri}^{\left(t\right)}\right),\\
 & \boldsymbol{v}_{B,pri}^{\left(t\right)}=\left(1/\boldsymbol{v}_{A,post}^{\left(t\right)}-1/\boldsymbol{v}_{A,pri}^{\left(t\right)}\right)^{-1}.
\end{aligned}
\label{eq:h_b_pri}
\end{equation}

\subsubsection{Module B}

\begin{figure}[t]
\begin{centering}
\textsf{\includegraphics[width=6.5cm]{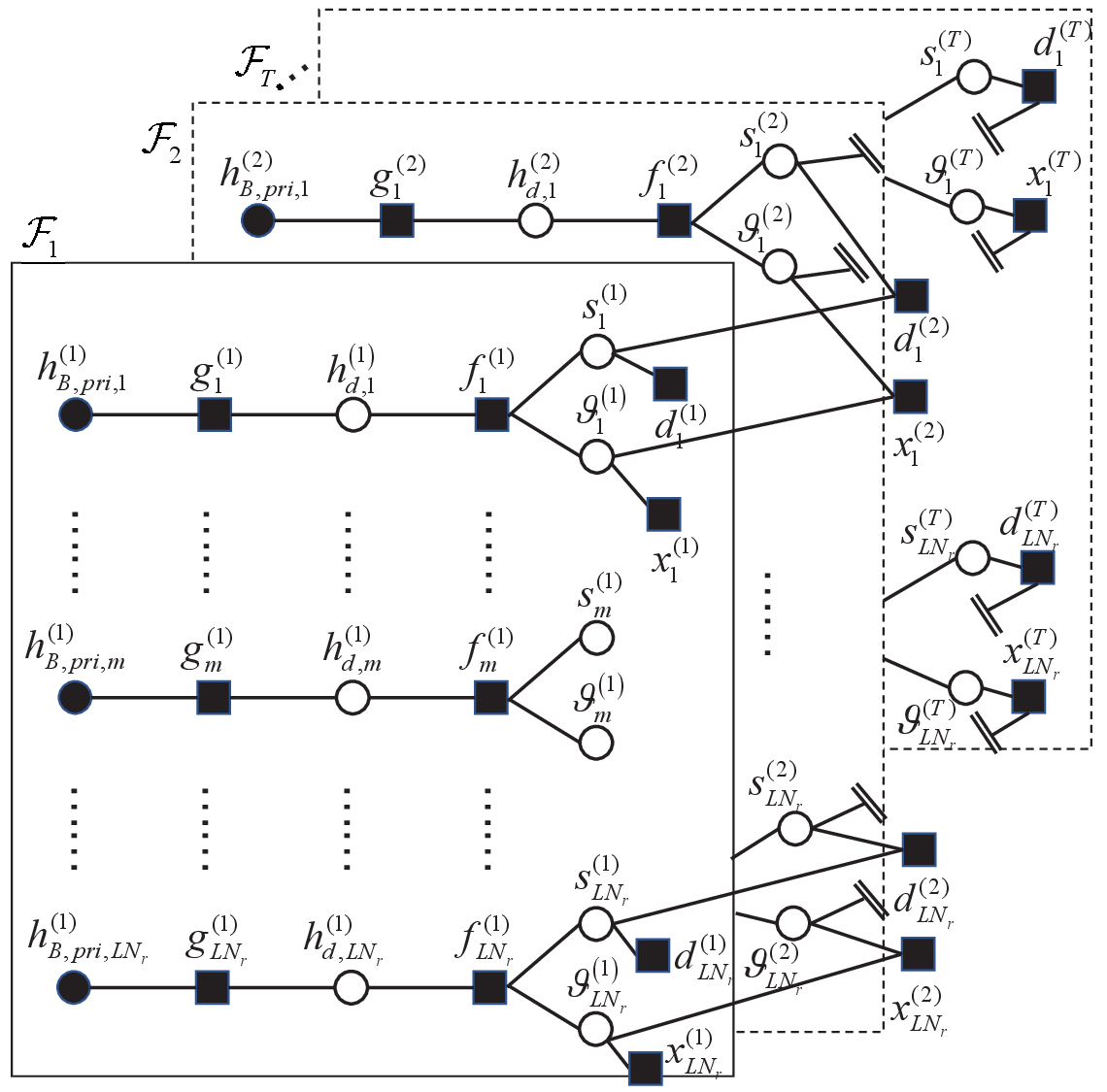}}
\par\end{centering}
\caption{\label{fig:Factor-graph-G}An illustration of factor graph $\mathcal{F}$.}
\end{figure}

We assume that $\boldsymbol{h}_{B,pri}^{(t)}$ is modeled as an AWGN
observation \cite{Turbo_CS,lian2019exploiting}:\vspace{-0.07in}
\begin{equation}
\boldsymbol{h}_{B,pri}^{(t)}=\boldsymbol{h}_{d}^{(t)}+\mathbf{z}^{(t)},\label{eq:ModuleBAWGN}
\end{equation}
where $\mathbf{z}^{(t)}\sim\mathcal{CN}(0,\textrm{diag}(\boldsymbol{v}_{B,pri}^{(t)}))$
is independent of $\boldsymbol{h}_{d}^{(t)}$. Under this assumption,
the factor graph (denoted as $\mathcal{F}$) of the joint probability
distribution $p(\boldsymbol{h}_{d}^{(T)},$$\boldsymbol{s}^{(T)},$$\boldsymbol{\vartheta}^{(T)},$$\boldsymbol{h}_{B,pri}^{(T)};$$\boldsymbol{\xi}^{\left(t\right)},$$\boldsymbol{\varDelta}^{(t)})$
is shown in Fig. \ref{fig:Factor-graph-G}, where the function expression
of each factor node is listed in Table \ref{tab:Factor-Distri-func}.
Based on (\ref{eq:ModuleBAWGN}), we combine the dynamic sparsity
prior information of $\boldsymbol{h}_{d}^{(t)}$ and the extrinsic
messages from Module A to calculate the posterior distributions $p(h_{d,m}^{(t)}\mid\boldsymbol{h}_{B,pri}^{(t)})$
by performing SPMP  over the factor graph $\mathcal{F}$. We now outline
the message passing procedure.
\begin{table}[t]
\centering{}{\small{}\caption{\label{tab:Factor-Distri-func}Factors and distributions forms in
Fig. \ref{fig:Factor-graph-G}.}
}{\footnotesize{}}%
\begin{tabular}{|>{\centering}m{2.5cm}|>{\centering}m{5.4cm}|}
\hline 
Factor & Distribution\tabularnewline
\hline 
\hline 
{\small{}$d_{m}^{(t)}(s_{m}^{(t)},s_{m}^{(t-1)})$} & {\small{}$\!\!\!\begin{array}{l}
\!\!p\left(s_{m}^{(t)}|s_{m}^{(t-1)}\right)\!\\
=\begin{cases}
\!\left(\rho_{10}\right)^{1-s_{m}^{(t)}}\!\left(1\!-\!\rho_{10}\right)^{s_{m}^{(t)}} & \!\!\!\!s_{m}^{(t-1)}\!=\!1\\
\!\left(\rho_{01}\right)^{s_{m}^{(t)}}\!\left(1\!-\!\rho_{01}\right)^{1-s_{m}^{(t)}} & \!\!\!\!s_{m}^{(t-1)}\!=\!0
\end{cases}
\end{array}$}\tabularnewline
\hline 
{\small{}$\begin{array}{c}
x_{m}^{(t)}(\vartheta_{m}^{(t)},\vartheta_{m}^{(t-1)})\end{array}$} & {\small{}$\begin{array}{l}
\!\!p\left(\vartheta_{m}^{\left(t\right)}|\vartheta_{m}^{\left(t-1\right)}\right)=\\
\mathcal{CN}\!\!\left(\vartheta_{m}^{\left(t\right)};\!\left(1\!-\beta_{\vartheta}\right)\vartheta_{m}^{\left(t-1\right)}\!+\!\beta_{\vartheta}\mu_{\vartheta},\beta_{\vartheta}^{2}\gamma_{\vartheta}\!\right)
\end{array}$}\tabularnewline
\hline 
{\small{}$f_{m}^{(t)}(\vartheta_{m}^{(t)},s_{m}^{(t)},h_{d,m}^{(t)})$} & \begin{lyxlist}{00.00.0000}
\item [{$\!\!\!\begin{array}{c}
p\left(h_{d,m}^{\left(t\right)}|\vartheta_{m}^{\left(t\right)},s_{m}^{\left(t\right)}\right)=\delta\left(h_{d,m}^{\left(t\right)}-\vartheta_{m}^{\left(t\right)}s_{m}^{\left(t\right)}\right)\end{array}$}]~
\end{lyxlist}
\tabularnewline
\hline 
{\small{}$g_{m}^{(t)}(h_{B,pri,m}^{(t)},h_{d,m}^{(t)})$} & {\small{}$p\left(h_{B,pri,m}^{\left(t\right)}|h_{d,m}^{\left(t\right)}\right)=\mathcal{CN}\left(h_{d,m}^{\left(t\right)};h_{B,pri,m}^{\left(t\right)},v_{B,pri,m}^{\left(t\right)}\right)$}\tabularnewline
\hline 
\end{tabular}
\end{table}

At $\mathcal{F}_{t}$, the message from variable node $h_{d,m}^{\left(t\right)}$
to factor node $f_{m}^{\left(t\right)}$ is $\nu_{h_{d,m}^{(t)}\rightarrow f_{m}^{(t)}}(h_{d,m}^{(t)})=\mathcal{CN}(h_{d,m}^{(t)};h_{B,pri,m}^{(t)},v_{B,pri,m}^{(t)})$
and the message passing is performed over the path $x_{m}^{(t)}\rightarrow\vartheta_{m}^{(t)}\rightarrow f_{m}^{(t)}$
and $d_{m}^{(t)}\rightarrow s_{m}^{(t)}\rightarrow f_{m}^{(t)}$,
respectively. Then, the marginal posterior distribution is given by\vspace{-0.15cm}
\begin{alignat}{1}
 & p\left(h_{d,m}^{(t)}|\boldsymbol{h}_{B,pri}^{\left(t\right)}\right)\!\propto\!\!\int_{\vartheta_{m}^{\left(t\right)}}\!\sum_{s_{m}^{(t)}}\!f_{m}^{(t)}\nu_{s_{m}^{(t)}\!\rightarrow f_{m}^{(t)}}\!\!\left(s_{m}^{(t)}\right)\nu_{\vartheta_{m}^{\left(t\right)}\!\rightarrow f_{m}^{(t)}}\!\!\left(\vartheta_{m}^{\left(t\right)}\right)\nonumber \\
 & \thinspace\thinspace\thinspace\thinspace\thinspace\thinspace\thinspace\thinspace\thinspace\thinspace\thinspace\thinspace\thinspace\thinspace\thinspace\thinspace\thinspace\thinspace\thinspace\thinspace\thinspace\thinspace\thinspace\thinspace\thinspace\thinspace\thinspace\thinspace\thinspace\thinspace\thinspace\thinspace\thinspace\thinspace\thinspace\thinspace\thinspace\thinspace\thinspace\thinspace\thinspace\thinspace\thinspace\thinspace\times\nu_{h_{d,m}^{\left(t\right)}\!\rightarrow f_{m}^{(t)}}\!\!\left(h_{d,m}^{\left(t\right)}\right).\label{eq:p_h}
\end{alignat}
Finally, the  extrinsic mean and variance are given by\vspace{-0.1cm}
\[
\boldsymbol{h}_{A,pri}^{\left(t\right)}=\boldsymbol{v}_{A,pri}^{\left(t\right)}\left(\boldsymbol{h}_{B,post}^{\left(t\right)}/\boldsymbol{v}_{B,post}^{\left(t\right)}-\boldsymbol{h}_{B,pri}^{\left(t\right)}/\boldsymbol{v}_{B,pri}^{\left(t\right)}\right),
\]
\begin{equation}
\begin{gathered}\boldsymbol{v}_{A,pri}^{\left(t\right)}=\left(1/\boldsymbol{v}_{B,post}^{\left(t\right)}-1/\boldsymbol{v}_{B,pri}^{\left(t\right)}\right)^{-1},\end{gathered}
\label{eq:h_a_pri}
\end{equation}
where $\boldsymbol{h}_{B,post}^{(t)}$ and $\boldsymbol{v}_{B,post}^{(t)}$
denote the posterior mean and variance corresponding to $p(h_{d,m}^{(t)}|\boldsymbol{h}_{B,pri}^{(t)})$.
After the convergence of the message passing over $\mathcal{F}_{t}$,
the message passing is performed across time $\mathcal{F}_{t}\rightarrow\mathcal{F}_{t+1}$
over the path $f_{m}^{\left(t\right)}\rightarrow s_{m}^{(t)}\rightarrow d_{m}^{\left(t+1\right)}$
and $f_{m}^{\left(t\right)}\rightarrow\vartheta_{m}^{\left(t\right)}\rightarrow x_{m}^{\left(t+1\right)}$,
and $\hat{p}(\boldsymbol{s}^{(t)})=\prod_{m=1}^{LN_{r}}\hat{p}(s_{m}^{(t)})$,
 $\hat{p}(\boldsymbol{\vartheta}^{(t)})=\prod_{m=1}^{LN_{r}}\hat{p}(\vartheta_{m}^{(t)})$,
where
\begin{equation}
\begin{aligned}\hat{p}\left(s_{m}^{(t)}\right) & \triangleq\nu_{s_{m}^{(t)}\rightarrow d_{m}^{\left(t+1\right)}}\left(s_{m}^{(t)}\right),\\
\hat{p}\left(\vartheta_{m}^{\left(t\right)}\right) & \triangleq\nu_{\vartheta_{m}^{\left(t\right)}\rightarrow x_{m}^{\left(t+1\right)}}\left(\vartheta_{m}^{\left(t\right)}\right).
\end{aligned}
\label{eq:ps_pvm}
\end{equation}
\vspace{-0.15in}

\subsection{M-Step\label{sec:M-step}}

In the M-Step, we construct a surrogate function at fixed point $\dot{\boldsymbol{\Xi}}^{(t)}$
for the objective function of the MAP problem in (\ref{eq:maxlikelyhood})
based on the MM method as \cite{Turbo_VBI,FDD2D} :\vspace{-0.15cm}
\begin{equation}
u(\boldsymbol{\Xi}^{(t)};\dot{\boldsymbol{\Xi}}^{(t)})\!\!=\!\!\int\!\!p(\boldsymbol{h}_{d}^{(t)}\!\mid\!\mathbf{y}^{(T)},\dot{\boldsymbol{\Xi}}^{(t)})\ln\frac{p(\boldsymbol{h}_{d}^{(t)}\!,\mathbf{y}^{(T)}\!,\boldsymbol{\Xi}^{(t)})}{p(\boldsymbol{h}_{d}^{(t)}\!\mid\!\mathbf{y}^{(T)},\dot{\boldsymbol{\Xi}}^{(t)})}d\boldsymbol{h}_{d}^{(t)},\label{eq:surrogatefunc}
\end{equation}
which satisfies basic properties
\[
u(\boldsymbol{\Xi}^{(t)};\dot{\boldsymbol{\Xi}}^{(t)})\leq\ln p(\boldsymbol{\Xi}^{(t)},\mathbf{y}^{(T)}),
\]
\begin{gather*}
u(\dot{\boldsymbol{\Xi}}^{(t)};\dot{\boldsymbol{\Xi}}^{(t)})=\ln p(\dot{\boldsymbol{\Xi}}^{(t)},\mathbf{y}^{(T)}),
\end{gather*}
\[
\left.\frac{\partial u(\boldsymbol{\Xi}^{(t)};\dot{\boldsymbol{\Xi}}^{(t)})}{\partial\boldsymbol{\Xi}^{(t)}}\right|_{\boldsymbol{\Xi}^{(t)}=\dot{\boldsymbol{\Xi}}^{(t)}}=\left.\frac{\partial\ln p(\boldsymbol{\Xi}^{(t)},\mathbf{y}^{(T)})}{\partial\boldsymbol{\Xi}^{(t)}}\right|_{\boldsymbol{\Xi}^{(t)}=\dot{\boldsymbol{\Xi}}^{(t)}},
\]
for $\forall\boldsymbol{\Xi}^{(t)}$. Then, we partition $\boldsymbol{\Xi}^{(t)}$
into $B=6$ blocks with $\Xi_{1}^{(t)}=\varepsilon^{(t)}$, $\Xi_{2}^{(t)}=\tau_{0}^{(t)}$,
$\boldsymbol{\Xi}_{3}^{(t)}=\boldsymbol{f}_{D}^{(t)}$, $\boldsymbol{\Xi}_{4}^{(t)}=\Delta\boldsymbol{\tau}^{(t)}$,
$\boldsymbol{\Xi}_{5}^{(t)}=\Delta\boldsymbol{\theta}^{(t)}$, $\boldsymbol{\Xi}_{6}^{(t)}=\Delta\boldsymbol{\phi}^{(t)}$
based on their distinct physical meaning, and alternatively update
$\Xi_{b}^{(t)}$ for $b=1,...,B$ as
\begin{align}
\boldsymbol{\Xi}_{b}^{(t)(j+1)} & =\mathrm{arg}\max_{\boldsymbol{\Xi}_{b}^{(t)}}u(\boldsymbol{\Xi}_{b}^{(t)},\boldsymbol{\Xi}_{-b}^{(t)(j)};\boldsymbol{\Xi}_{b}^{(t)(j)},\boldsymbol{\Xi}_{-b}^{(t)(j)}),\label{eq:update PN}
\end{align}
where $\boldsymbol{\Xi}_{-b}^{(t)(j)}=(\boldsymbol{\Xi}_{1}^{(t)(j+1)},...,\boldsymbol{\Xi}_{b-1}^{(t)(j+1)},\boldsymbol{\Xi}_{b+1}^{(t)(j)},...,\boldsymbol{\Xi}_{B}^{(t)(j)})$,
and $j$ stands for the $j$-th iteration. We can obtain a closed-form
solution of (\ref{eq:update PN}) for $\varepsilon^{(t)}$ as $\hat{\varepsilon}^{(t)}$$=$$\angle((\boldsymbol{\Phi}^{(t)}\boldsymbol{\mu}_{d}^{(t)})^{H}\mathbf{y}^{(t)})$,
where $\boldsymbol{\mu}_{d}^{(t)}$ is the posterior mean of $\hat{p}(\boldsymbol{h}_{d}^{(t)}|\mathbf{y}^{(T)};\boldsymbol{\Xi}^{(t)})$
estimated in the E-Step. However, the surrogate functions for other
variables are non-convex and it is difficult to find their optimal
solutions. Therefore, we use a fixed stepsize one-step gradient update
as in \cite{FDD2D}, i.e.,\vspace{-0.1cm}
{\small{}
\begin{align}
\boldsymbol{\Xi}_{b}^{(t)(j+1)}=\boldsymbol{\Xi}_{b}^{(t)(j)}+\frac{\gamma_{b}}{50}\cdot\textrm{sign}(\boldsymbol{\zeta}_{b}^{(j)}), & b=1,...,B,\label{eq:ugrad}
\end{align}
}where $\gamma_{b}$ stands for the grid interval, $\boldsymbol{\zeta}_{b}^{(j)}$
denotes the gradient of the objective function (\ref{eq:surrogatefunc})
with respect to $\boldsymbol{\Xi}_{b}^{(t)}$, and sign(·) stands
for the signum function. The convergence of this MM based algorithm
to a stationary point is guaranteed \cite[Theorem 1]{Turbo_VBI}.
Our proposed tracking scheme exploits  prior information  in the M-Step,
i.e., $\hat{\boldsymbol{f}}_{D}^{(t-1)},\hat{\boldsymbol{\Delta}}^{(t-1)}$,
based on the probability model (\ref{eq:imperfections})-(\ref{eq:offgrid-1})
and has been well initialized. Hence, a good solution can always be
found while the original EM method may easily trap into ``bad''
local optimums.

Finally, the overall channel extrapolation scheme in tracking stage
at time $t$ is summarized in Algorithm \ref{alg:Channel-tracking-algorithm}.
\begin{algorithm}[t]
\caption{\label{alg:Channel-tracking-algorithm}Channel extrapolation scheme
in tracking stage}

\textbf{Input:} $\mathbf{Y}^{(t)},\hat{\boldsymbol{f}}_{D}^{(t-1)},\hat{\boldsymbol{\Delta}}^{(t-1)}$,
EM iteration number $I_{EM}$.

\textbf{Output:} The recovered full-band CFR $\hat{\mathbf{H}}^{(t)}$,
$\boldsymbol{\hat{\xi}}^{(t)},\hat{\boldsymbol{\Delta}}^{(t)}$. 

\begin{algorithmic}[1]

\STATE\textbf{Initialization: $\boldsymbol{\hat{f}}_{D}^{(t)}=\boldsymbol{\hat{f}}_{D}^{(t-1)},\boldsymbol{\hat{\Delta}}^{(t)}=\boldsymbol{\hat{\Delta}}^{(t-1)}$.}

\FOR{${\color{blue}{\color{black}j=1,\cdots,I_{EM}}}$}

\STATE\textbf{E-Step:}

\WHILE{not converge}

\STATE \textbf{\%Module A: LMMSE Estimator}

\STATE Calculate $\boldsymbol{V}_{A,post}^{(t)},\boldsymbol{h}_{A,post}^{(t)},\boldsymbol{h}_{B,pri}^{(t)},\boldsymbol{v}_{B,pri}^{(t)}$
in (\ref{eq:Va_post})-(\ref{eq:h_b_pri}).

\STATE \textbf{\%Module B: Sparsity Combiner}

\STATE Perform message passing over graph $\mathcal{F}_{t}$.

\STATE Calculate $p(h_{d,m}^{(t)}|\boldsymbol{h}_{B,pri}^{(t)})$
in (\ref{eq:p_h}).

\STATE Update $\boldsymbol{h}_{A,pri}^{(t)}$ and $\boldsymbol{v}_{A,pri}^{(t)}$in
(\ref{eq:h_a_pri}).

\ENDWHILE

\STATE\textbf{M-Step}

\STATE Construct surrogate function in (\ref{eq:surrogatefunc}).

\STATE Update  $\boldsymbol{\Xi}_{b}^{(t)(j+1)},\forall b,$ in (\ref{eq:update PN}).

\ENDFOR

\STATE Perform message passing $\mathcal{F}_{t}\rightarrow\mathcal{F}_{t+1}$. 

\STATE Calculate the recovered full-band CFR $\hat{\mathbf{H}}^{(t)}$
based on the estimated results $\hat{p}(\boldsymbol{h}_{d}^{(t)}|\mathbf{y}^{(T)})$
and $\hat{\boldsymbol{\Xi}}^{(t)}$. 

\STATE Pass the estimate $\hat{\boldsymbol{f}}_{D}^{(t)},\hat{\boldsymbol{\Delta}}^{(t)}$,
and $\hat{p}(\boldsymbol{s}^{(t)})$, $\hat{p}(\boldsymbol{\vartheta}^{(t)})$
to the next time $\left(t+1\right)$.

\end{algorithmic}
\end{algorithm}
\vspace{-0.1in}

\subsection{Computational Complexity Analysis}

The computational complexity of the channel tracking scheme in E-Step
is dominated by the inverse operation in (\ref{eq:Va_post}), which
is $\mathcal{O}\left(N_{r}^{3}L^{3}\right)$, matrix multiplication
in (\ref{eq:Va_post}) to calculate $\boldsymbol{V}_{A,post}^{\left(t\right)}$
and $\boldsymbol{h}_{A,post}^{\left(t\right)}$, which is $\mathcal{O}\left(N_{r}^{3}L^{2}P\right)$
and $\mathcal{O}\left(N_{r}^{2}LP\right)$. Besides, the main computational
complexity in M-Step is $\mathcal{O}\left(N_{r}^{3}L^{2}P\right)$
per iteration. Note that $L$ can be small in our problem since we
adopt the off-grid adjustment strategy, which do not require a dense
grid to guarantee the delay estimation accuracy.\vspace{-0.1in}

\section{Simulation Results\label{sec:Simulation-Results}}

In this section, we provide numerical results to evaluate the channel
estimation performance of the proposed scheme. The MIMO-OFDM system
is equipped with carrier frequency 3.5 GHz, the bandwidth $B$ = 60
MHz, and the subcarrier spacing $f_{s}=60$ KHz. The BS is equipped
with $N_{r}=64$ ($N_{x}=N_{y}=8$) antennas and the user moves with
speed 3 km/h. The bandwidth of each BWP is 30 MHz (15 MHz) and the
number of the SRS sequence is $P=250$ (125) for $h_{p}=2$ (4).\textcolor{blue}{{}
}The $\textrm{SNR}=15$ dB and the delay grid size $L=26$. The CFR
samples are generated by the QuaDRiGa toolbox \cite{Quadriga} according
to the 3D-UMa NLOS model defined by 3GPP R16 specifications \cite{3gpp_Rel16}
and the performance result of the algorithms is averaged over 500
noise realizations. We choose normalized mean square error (NMSE)
as the performance metric to evaluate the extrapolation performance
of various algorithms, which is defined as $\textrm{NMSE}=\frac{\left\Vert \hat{\mathbf{H}}^{(t)}-\mathbf{H}^{(t)}\right\Vert _{F}^{2}}{\left\Vert \mathbf{H}^{(t)}\right\Vert _{F}^{2}}$.

For comparison, we consider the following three benchmark schemes
and  use the same hopping SRSs pattern for all schemes for fairness.
\begin{itemize}
\item \textbf{Baseline 1} : We employ the TST-MUSIC algorithm to perform
channel extrapolation at each timeslot independently \cite{TST_MUSIC}.
\item \textbf{Baseline 2} : We employ the OAMP based channel tracking algorithm
to perform channel extrapolation \cite{lian2019exploiting}. Specifically,
the channel tracking is performed at each BWP independently without
frequency extrapolation and the BWP owning received SRSs can be estimated.
Then, to achieve full-band channel estimation, a simple extrapolation
in time-domain is employed: For each BWP, the rest channel (i.e.,
the white parts of channel in Fig. \ref{fig:baseline}) estimation
is seen as equivalent to the channel estimation results at the latest
time, e.g., as shown in Fig. \ref{fig:baseline}, the channel estimation
results at the blocks with the same linestyle are equal.
\item \textbf{Baseline 3} We employ the proposed algorithm but without performing
imperfection factors compensation and off-grid update, i.e., using
TST-MUSIC algorithm in the channel estimation stage and proposed tracking
algorithm without M-Step in the channel tracking stage. 
\begin{figure}[t]
\begin{centering}
\textsf{\includegraphics[width=5cm]{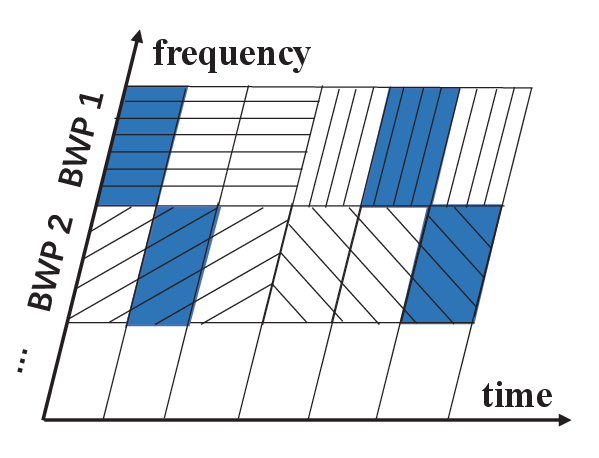}}
\par\end{centering}
\caption{\label{fig:baseline}An illustration of Baseline 2.}
\end{figure}
\end{itemize}

\begin{figure}[t]
\centering{}\textcolor{blue}{\includegraphics[width=8cm]{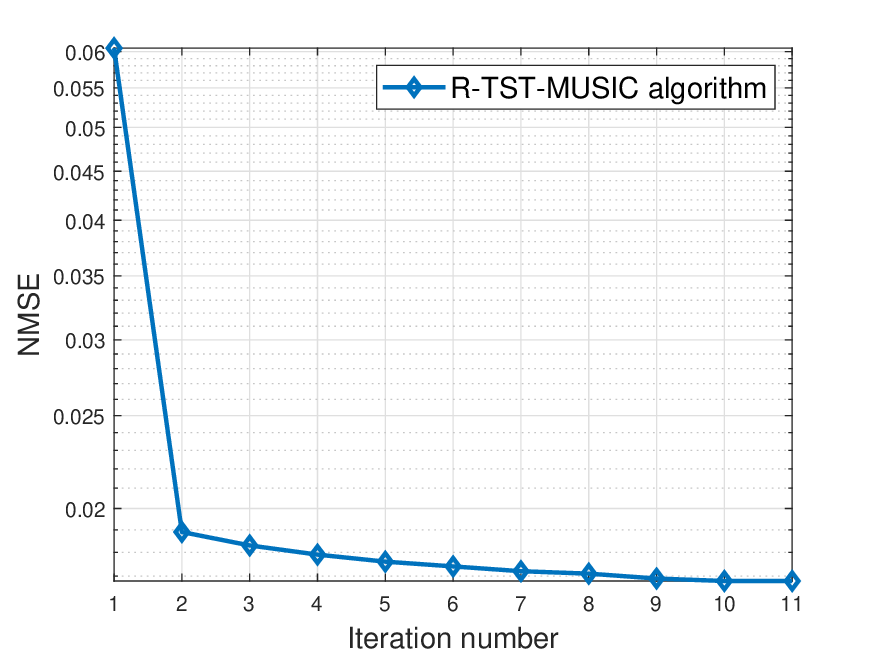}\caption{\label{fig:Convergence-behavior-of}Convergence behavior of the R-TST-MUSIC
algorithm.}
}
\end{figure}

We first illustrate the convergence behavior of the proposed R-TST-MUSIC
algorithm. As illustrated in Fig. \ref{fig:Convergence-behavior-of},
R-TST-MUSIC converges within 10 iterations (up to a small convergence
error).\textcolor{blue}{}

\begin{figure}[t]
\begin{centering}
\begin{minipage}[t]{0.45\textwidth}%
\begin{center}
\subfloat[]{\begin{centering}
\includegraphics[width=80mm]{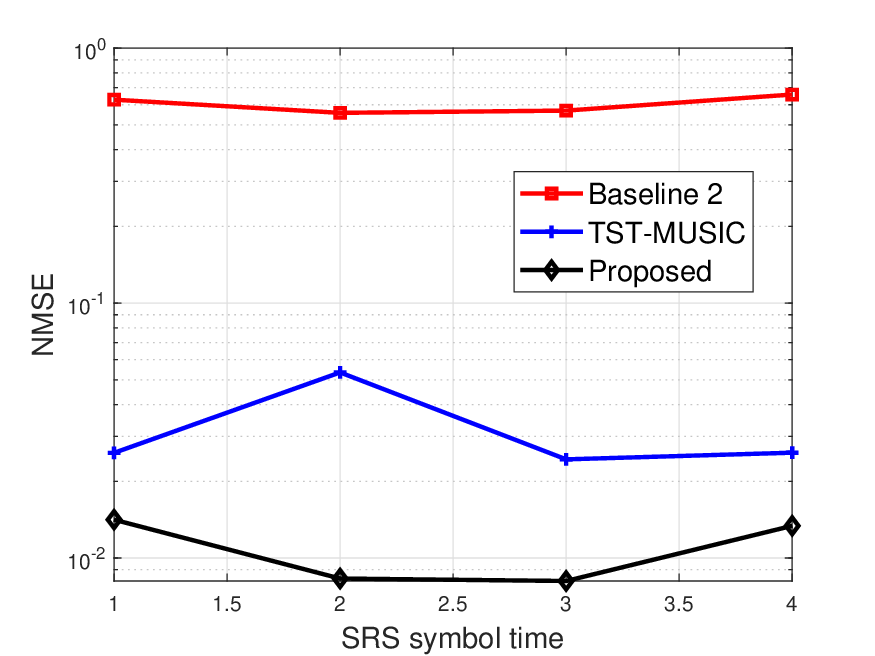}
\par\end{centering}
}
\par\end{center}%
\end{minipage}\hfill{}%
\begin{minipage}[t]{0.45\textwidth}%
\begin{center}
\subfloat[]{\begin{centering}
\includegraphics[width=80mm]{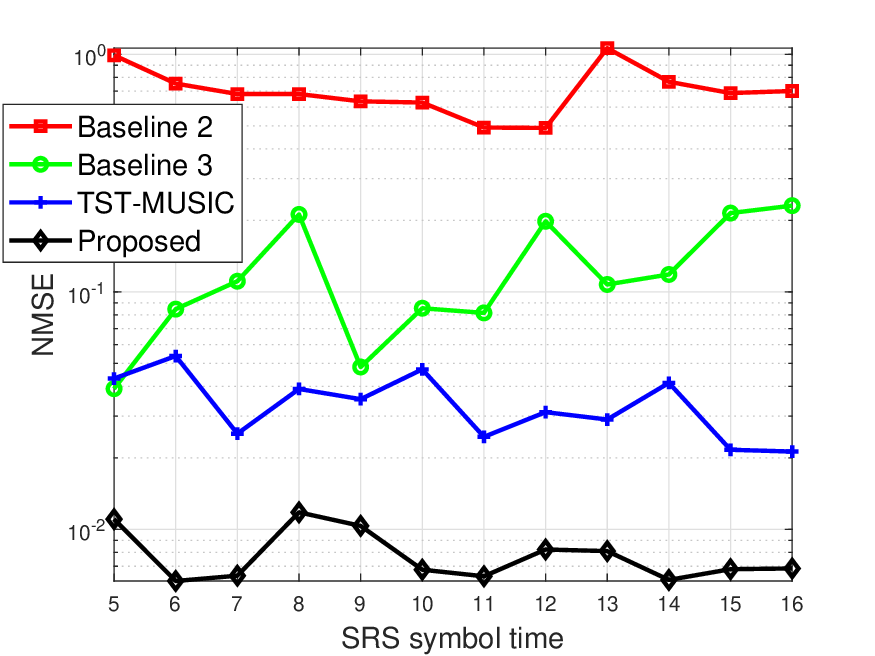}
\par\end{centering}
}
\par\end{center}%
\end{minipage}
\par\end{centering}
\medskip{}

\centering{}\caption{\label{fig:fullband_time}NMSE of the full-band channel versus time
for $h_{p}=4$: (a) Channel estimation stage; (b) Channel tracking
stage.}
\end{figure}

In Fig. \ref{fig:fullband_time}, we present the NMSE performance
of the full-band channel versus time for various algorithms in both
channel estimation and tracking stage. First, it can be seen that
our proposed R-TST-MUSIC algorithm reaps a significant performance
gain compared with original TST-MUSIC algorithm. Second, the extrapolation
schemes (i.e., the proposed scheme and Baseline 3) achieve a better
performance than the traditional channel tracking algorithm, i.e.,
Baseline 2. On one hand, it is mainly because Baseline 2 do not have
a real extrapolation ability, i.e., without extrapolation in frequency
domain, and the extrapolation in time domain has approximation error.
Besides, since the channel for each BWP is estimated independently,
it cannot fully exploit all observation information at different BWPs
to improve the channel parameter estimation accuracy. On the other
hand, as compared to Baseline 2, our proposed two-stage channel extrapolation
scheme performs a meticulously designed multi-timeslot based channel
estimation initially to provide a better initial value for doing channel
tracking in the second stage. Finally, we observe that the proposed
scheme outperforms Baseline 3 and TST-MUSIC algorithm, which demonstrates
the necessity of the imperfection factors compensation and employing
the off-grid channel model.

\begin{figure}[t]
\begin{centering}
\includegraphics[width=80mm]{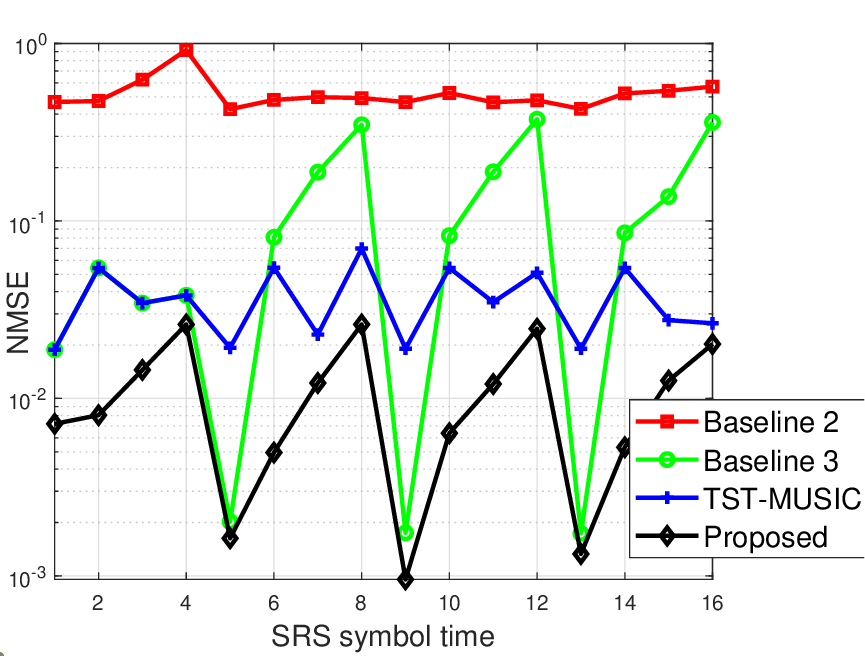}
\par\end{centering}
\centering{}\caption{\label{fig:BWP_time}NMSE of the first BWP channel versus time for
$h_{p}=4$.}
\end{figure}

In Fig. \ref{fig:BWP_time}, the NMSE performance of the first BWP
channel is illustrated as a function of SRS symbol time. As can be
seen, our proposed scheme outperforms other schemes for all times.
Besides, the curves of Baseline 3 and the proposed scheme oscillate
with the period $h_{p}$. It is reasonable since the position of SRSs
in frequency domain undergoes periodic changes, and within a cycle,
the first BWP becomes increasingly distant from the position of the
SRSs in frequency domain as time increases, resulting in an increasing
extrapolation distance. Furthermore, Baseline 3 exhibits the most
intense oscillations, which indicates that the imperfection parameters
will seriously affect the algorithm's extrapolation ability if they
are not well compensated for.

\begin{figure}[t]
\centering{}\includegraphics[width=8cm]{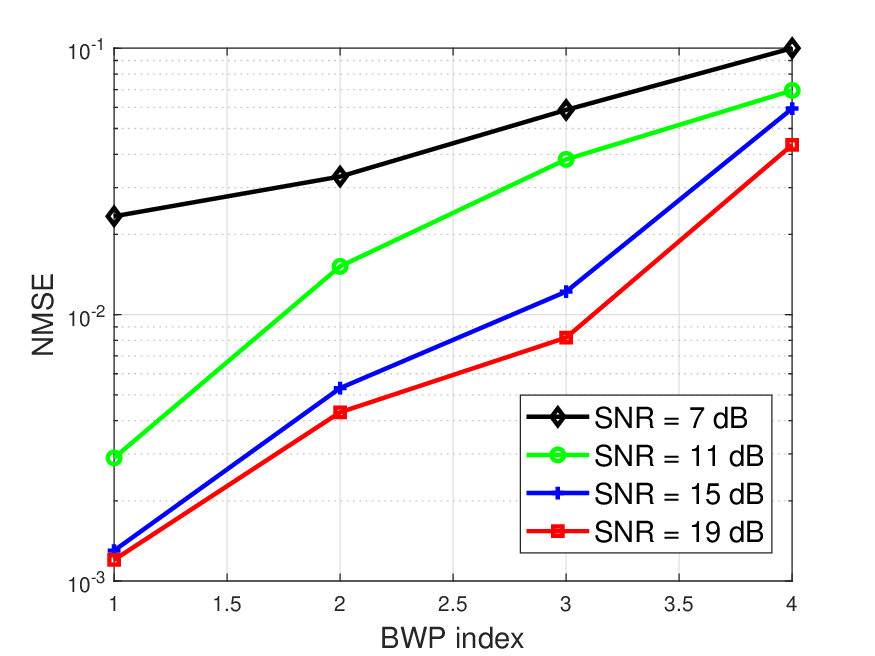}\textcolor{blue}{\caption{\label{fig:BWP_index}NMSE of different BWPs channel for $h_{p}=4$.}
}
\end{figure}

Then, we investigate the extrapolation performance of our proposed
scheme focusing on the $1+h_{p}n,n=0,...,N_{s}-1,$-th SRS symbol
time, at all of which the SRSs locate in the first BWP. Fig. \ref{fig:BWP_index}
depicts the NMSE performance of different BWPs for various SNRs with
$h_{p}=4$, where the NMSE is averaged over $N_{s}=15$ SRS symbol
times. It is observed that the NMSE increases with the BWP index due
to the increased extrapolation range.

\begin{figure}[t]
\begin{centering}
\begin{minipage}[t]{0.45\textwidth}%
\begin{center}
\subfloat[]{\begin{centering}
\includegraphics[width=80mm]{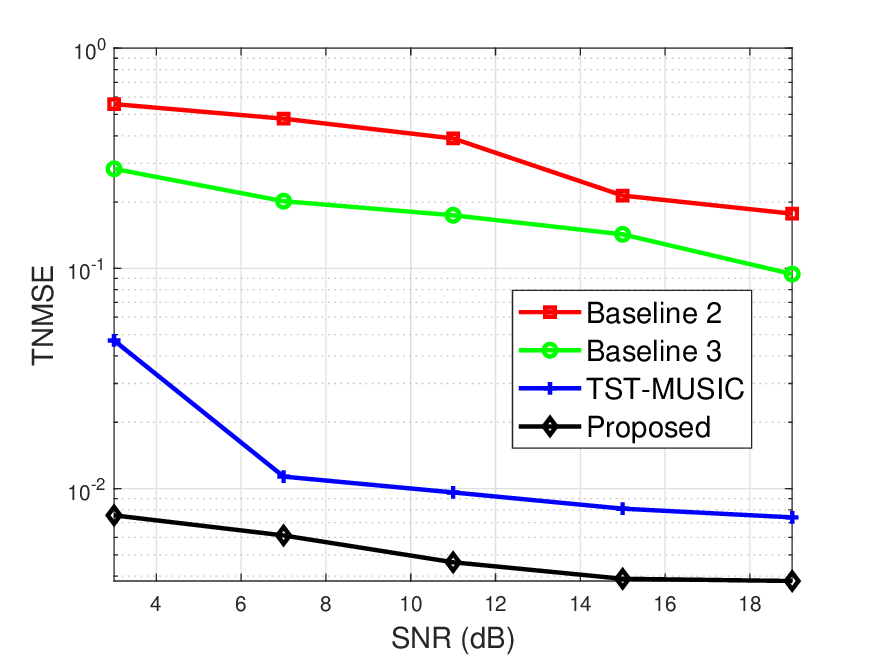}
\par\end{centering}
}
\par\end{center}%
\end{minipage}\hfill{}%
\begin{minipage}[t]{0.45\textwidth}%
\begin{center}
\subfloat[\label{fig:Convergeb}]{\begin{centering}
\includegraphics[width=80mm]{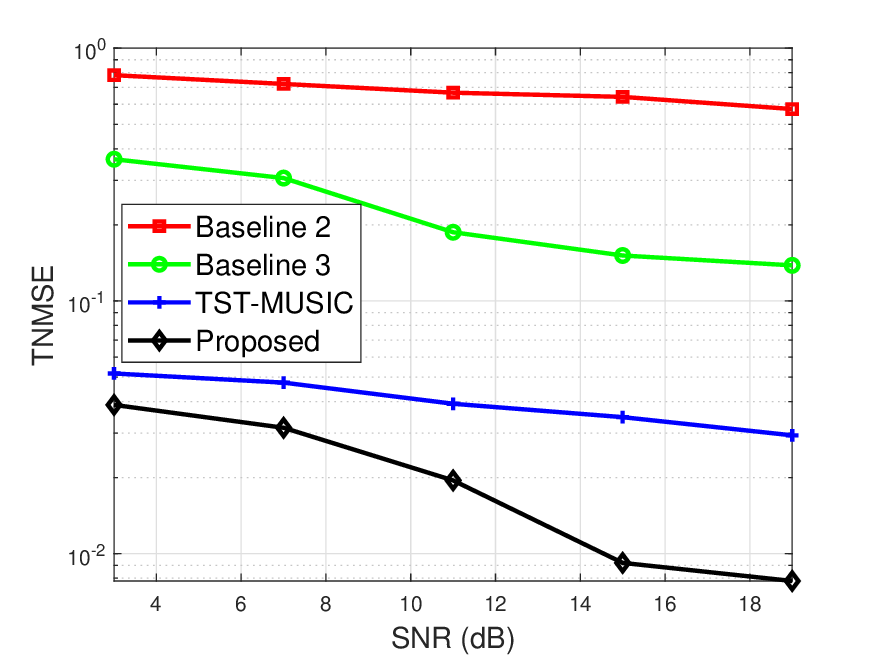}
\par\end{centering}
}
\par\end{center}%
\end{minipage}
\par\end{centering}
\medskip{}

\centering{}\caption{\label{fig:TNMSE_SNR}TNMSE versus SNR for: (a) $h_{p}=2$; (b) $h_{p}=4$.}
\end{figure}

In Fig. \ref{fig:TNMSE_SNR}, we investigate the time-averaged NMSE
(TNMSE) performance versus SNR for $h_{p}=2$ and $h_{p}=4$, respectively,
where $\textrm{TNMSE}=\frac{1}{T}\sum_{t=1}^{T}\frac{\left\Vert \hat{\mathbf{H}}^{(t)}-\mathbf{H}^{(t)}\right\Vert _{F}^{2}}{\left\Vert \mathbf{H}^{(t)}\right\Vert _{F}^{2}}$
with $T=60$. As can be seen, the TNMSE of all schemes decreases as
SNR increases. Besides, our proposed scheme achieves significant performance
gain compared with baselines for both $h_{p}=2$ and $h_{p}=4$. Furthermore,
the algorithms have better performance in the case of $h_{p}=2$ than
$h_{p}=4$ due to a relatively narrower extrapolation range.

\vspace{-0.1in}

\section{Conclusion\label{sec:Conclusion}}

In this paper, we proposed a two-stage 2D channel extrapolation scheme
compatible with TDD massive MIMO 5G NR systems. We constructed a new
received signal model for 2D channel extrapolation in the presence
of imperfection factors based on a hopping pilot pattern. Then, in
the channel estimation stage, we proposed a novel MBMT-HRPE scheme
to compensate for the imperfection factors and achieve high-accuracy
channel extrapolation. To avoid frequent multi-timeslot based channel
estimation, we adopted a channel tracking scheme in the second stage.
 Finally, simulation results validated the effectiveness of our proposed
channel extrapolation scheme. 

\vspace{-0.1in}

\bibliographystyle{IEEEtran}
\bibliography{channel_extrapolation}

\end{document}